\documentclass{article}

\usepackage{fullpage}
\usepackage{lmodern}
\usepackage[utf8]{inputenc}
\usepackage[T1]{fontenc}
\usepackage[USenglish]{babel}

\usepackage{amsmath, amssymb, amsthm}

\usepackage{booktabs}

\usepackage{algorithm}
\usepackage[noend]{algpseudocode}

\algrenewtext{While}[1]{\algorithmicwhile\ #1\textbf{:}}
\algrenewtext{For}[1]{\algorithmicfor\ #1\textbf{:}}
\algrenewtext{ForAll}[1]{\algorithmicforall\ #1\textbf{:}}
\algrenewtext{If}[1]{\algorithmicif\ #1\textbf{:}}
\algrenewtext{ElsIf}[1]{\algorithmicelse\ \algorithmicif\ #1\textbf{:}}
\algrenewcommand\algorithmicrequire{\textbf{Input:}}
\algrenewcommand\algorithmicensure{\textbf{Output:}}

\algnewcommand{\InlineFor}[1]{\State \algorithmicfor\ #1\textbf{:}}
\algnewcommand{\InlineForAll}[1]{\State \algorithmicforall\ #1\textbf{:}}
\algnewcommand{\InlineIf}[1]{\State \algorithmicif\ #1\textbf{:}}
\algnewcommand{\InlineElse}[1]{\State \algorithmicelse\textbf{:}}
\algnewcommand{\algorithmiclet}{\textbf{let}}
\algnewcommand{\Let}{\algorithmiclet\ }

\algnewcommand{\algorithmicoutput}{\textbf{output}}
\algnewcommand{\Output}{\algorithmicoutput\ }

\algnewcommand{\algorithmiccontinue}{\textbf{continue}}
\algnewcommand{\Continue}{\algorithmiccontinue\ }

\usepackage{enumitem}

\usepackage[round]{natbib}
\usepackage[%
  pdftitle={The Time-Dependent Traveling Salesman Problem with Loose Time Windows},%
  pdfauthor={Francisco J.\ Soulignac},%
  colorlinks=true,%
  linkcolor=blue,%
  citecolor=blue,%
  urlcolor=blue]{hyperref}
\usepackage{doi}

\usepackage{tikz}
\usetikzlibrary{graphs, quotes, graphs.standard}
\usepackage{float}
\usepackage{caption}
\usepackage{subcaption}

\newcommand{\TSP}{\mbox{TSP}}
\newcommand{\TSPTW}{\mbox{TSPTW}}
\newcommand{\TDTSPTW}{\mbox{TDTSPTW}}
\newcommand{\TDTSP}{\mbox{TDTSP}}
\newcommand{\Network}{D}
\newcommand{\Memory}{M}

\newcommand{\Horizon}{T}
\newcommand{\Release}{a}
\newcommand{\Deadline}{b}
\newcommand{\CustCount}{n}
\newcommand{\Position}{p}
\newcommand{\OtherRoute}{q}
\newcommand{\Route}{r}
\newcommand{\Time}{t}
\newcommand{\Vertex}{v}
\newcommand{\Wertex}{w}
\newcommand{\DecisionVar}{y}
\newcommand{\Zertex}{z}
\newcommand{\Repetitions}{\alpha}
\newcommand{\ArrivalTime}{\delta}
\newcommand{\RedArrivalTime}{\overline{\ArrivalTime}}
\newcommand{\Profit}{\sigma}
\newcommand{\TotalProfit}{\Sigma}
\newcommand{\TravelTime}{\tau}
\newcommand{\BackTravelTime}{\TravelTime^{\rm b}}
\newcommand{\DepartureTime}{\ArrivalTime^{\rm b}}
\newcommand{\RedDepartureTime}{\RedArrivalTime\vphantom{\ArrivalTime}^{\rm b}}
\newcommand{\RedCost}{\RedArrivalTime\vphantom{\ArrivalTime}^{\rm cb}}
\newcommand{\Range}[1]{[\![#1]\!]}
\newcommand{\LowerBound}{lb}
\newcommand{\PricingLowerBound}{\overline{lb}}
\newcommand{\Sequence}[1]{\langle#1\rangle}
\newcommand{\RouteSet}{\mathcal{R}}
\newcommand{\Tree}{\mathcal{F}}
\newcommand{\BackTree}{\mathcal{B}}
\newcommand{\Label}{\ell}
\newcommand{\Mabel}{m}
\newcommand{\RouteUniverse}{\Omega}
\newcommand{\BackRouteUniverse}{\RouteUniverse^{\rm b}}
\newcommand{\LabelSet}{\mathcal{L}}
\newcommand{\BackLabelSet}{\LabelSet^{\rm b}}
\newcommand{\SparsificationCost}{\rho}
\newcommand{\Stage}{S}
\newcommand{\RMPBound}{d^*}
\newcommand{\MaxPricingIterations}{\iota}

\newcommand{\CompletionBound}{cb}
\newcommand{\Adamo}{\textsc{Ada20}}
\newcommand{\Arigliano}{\textsc{Ari19}}
\newcommand{\Rifki}{\textsc{Rif20}}
\newcommand{\RifkiNoTW}{\textsc{Rif20}$^{[\Horizon]}$}
\newcommand{\RifkiConst}{\textsc{Rif20}$^c$}
\newcommand{\Lera}{\textsc{Ler22}}
\newcommand{\Fontaine}{\textsc{Fon23}}
\newcommand{\TW}{\beta}
\newcommand{\Pesant}{\textsc{Pes98}}
\newcommand{\PotvinBengio}{\textsc{Pot96}}
\newcommand{\Vu}{\textsc{Vu20}}
\newcommand{\BP}{\textsc{B\&P}}
\newcommand{\Base}{\textsc{Base}}
\newcommand{\Sparse}{\textsc{NoPH}}
\newcommand{\PHOnly}{\textsc{NoSP}}

\DeclareMathOperator{\Prev}{last}
\DeclareMathOperator{\Next}{next}
\DeclareMathOperator{\Master}{MP}

\newtheorem{observation}{Observation}
\newtheorem{proposition}{Proposition}
\newtheorem{domrule}{Rule}
\newtheorem{remark}{Remark}

\newcommand{\CPP}{C\nolinebreak\hspace{-.05em}\raisebox{.2ex}{\small\bf +}\nolinebreak\hspace{-.10em}\raisebox{.2ex}{\small\bf +}20}

\title{The Time-Dependent Traveling Salesman Problem with Loose Time Windows}

\author{Francisco J.\ Soulignac\\\normalsize \texttt{francisco.soulignac@unq.edu.ar}}

\date{\normalsize Universidad Nacional de Quilmes. Departamento de Ciencia y Tecnología. Bernal, Buenos Aires, Argentina.\\ CONICET-Universidad de Buenos Aires. Instituto de Investigación en Ciencias de la Computación (ICC). Buenos Aires, Argentina.}

\begin{document}

\maketitle

\begin{abstract}

The time-dependent traveling salesman problem with time windows (\TDTSPTW{}) generalizes the well-known traveling salesman problem with time windows by accounting the effects of congestion on travel times. In this paper, we develop an exact framework for the \TDTSPTW{} with a makespan objective that extends the range of instances solvable to optimality under loose time windows while remaining effective across all levels of time-window tightness. Our framework relies on a dynamic-programming labeling algorithm and combines column generation, ng-memory augmentation, and exact search, using completion bounds for state-space sparsification, variable fixing, and exact search pruning. Embedded within a branch-and-price method, the framework solves all instances with up to 45 customers in a benchmark comprising more than 10,000 instances, including all instances without time windows with up to 50 customers.

~

\noindent\textbf{Keywords:} traveling salesman problem, time-dependent travel times, loose time windows, dynamic programming, column generation
\end{abstract}

\section{Introduction}

The Traveling Salesman Problem with Time Windows (\TSPTW{}) seeks a tour that visits a set of customers within given time windows, assuming fixed travel times. In practice, however, congestion may cause travel times to vary over the planning horizon. The Time-Dependent \TSPTW{} (\TDTSPTW{}), introduced by \cite{MalandrakiDaskinTS1992}, accounts for this variation by allowing travel times to depend on the departure time. We consider piecewise-linear travel-time functions satisfying the FIFO property, whereby departing later on an arc cannot result in an earlier arrival. Our objective is to minimize tour completion time (makespan) from a fixed departure time of zero. For broader discussions of time-dependent routing, including FIFO and alternative travel-time models and objective functions, see \cite{IchouaGendreauPotvinEJOR2003,GendreauGhianiGuerrieroCOR2015,Fontaine2024,AdamoGendreauGhianiGuerrieroEJOR2024}.

Research on the \TSPTW{} began more than forty years ago with the foundational works of \cite{ChristofidesMingozziToth1981} and \cite{Baker1983}. Since then, a large and still evolving body of literature has developed, encompassing exact and heuristic approaches \citep[see][for an updated literature review]{Pralet2023,SoulignacCOR2026}. The most efficient exact methods currently rely on dynamic programming, also for related objectives such as travel cost and duration minimization. These include anytime informed-search methods \citep{FontaineDibangoyeSolnon2023,KuroiwaBeck2023,SoulignacCOR2026}, decremental state-space relaxation frameworks based on ng-memory \citep{BaldacciMingozziRoberti2012,TilkIrnich2017,Lera-RomeroMirandaBrontSoulignac2022}, and general-purpose multivalued decision diagrams, which have achieved competitive results on \TSPTW{} instances with tight time windows \citep{GillardCoppeSchausCire2021,RudichCappartRousseau2023,CoppeGillardSchaus2024,TardivoMichelHoeve2026}. Dynamic programming is also used in heuristic approaches, such as the iterated large-neighborhood search of \cite{Pralet2023}, which can prove optimality on sufficiently constrained instances.

The \TDTSPTW{} has attracted a smaller but steadily growing body of research since its introduction by \cite{MalandrakiDaskinTS1992}. Exact approaches include constraint programming \citep{MelgarejoLaborieSolnon2015}, branch-and-bound and branch-and-cut \citep{CordeauGhianiGuerrieroTS2014,MonteroMendez-DiazMirandaBrontCO2017,AriglianoCalogiuriGhianiGuerrieroN2018,AriglianoGhianiGriecoGuerrieroPlana2019,AdamoGhianiGuerrieroCOR2020}, dynamic discretization discovery \citep{VuHewittBolandSavelsberghTS2020}, and dynamic programming \citep{Lera-RomeroMirandaBrontSoulignac2022,FontaineDibangoyeSolnon2023}. The dynamic programming approaches of \cite{FontaineDibangoyeSolnon2023} and \cite{Lera-RomeroMirandaBrontSoulignac2022} are currently the most efficient methods for the \TDTSPTW{}, with the former performing particularly well on tightly constrained instances and the latter becoming comparatively more effective as the windows loosen. Together with \cite{SoulignacCOR2026}, these approaches define the state of the art for the \TSPTW{}, with \cite{SoulignacCOR2026} being particularly effective on the most tightly constrained instances.

The width of the time windows is a key factor in the efficiency of dynamic programming approaches for the \TSPTW{} and its time-dependent variants, and can be more important than the number of customers. Tight time windows restrict the set of feasible tours and thus the search space, whereas looser windows allow substantially larger search spaces. This behavior has been observed in both the \TSPTW{} \citep{DumasDesrosiersGelinasSolomon1995,SoulignacCOR2026} and the \TDTSPTW{} \citep{Lera-RomeroMirandaBrontSoulignac2022,FontaineDibangoyeSolnon2023}. Further evidence is provided by \cite{RifkiSolnon2025}, who show that the ratio between time-window width and the planning horizon is a better predictor of computational effort than the number of customers for the exact approach of \cite{FontaineDibangoyeSolnon2023}.

For the \TSPTW{}, computational evaluations commonly rely exclusively on the benchmark collection maintained by \cite{Lopez-IbanezBlum2023}. Recently, \cite{SoulignacCOR2026} reported that a straightforward exact algorithm solves all classical benchmark instances with 50 or more customers within seconds, yet struggles on newer instances introduced by \cite{Fontaine2024} with fewer than 30 customers and loose time windows. These results suggest that loose time-window regimes are underrepresented in the classical benchmark collection, highlighting the importance of evaluating exact methods across a broad spectrum of temporal constraints. Interestingly, \cite{Fontaine2024} adapts time-dependent instances of \cite{RifkiChiabautSolnon2020}, systematically varying their time-window tightness, following a strategy introduced by \cite{AriglianoGhianiGriecoGuerrieroPlana2019} for the \TDTSPTW{}. The state-of-the-art comparison above draws on the computational experiments of \cite{Fontaine2024}, which evaluate leading exact solvers across these regimes.

As time windows become looser, the \TDTSPTW{} gradually approaches the Time-Dependent Traveling Salesman Problem (\TDTSP{}), in which no time-window constraints are imposed. An intermediate case arises when only deadlines are present, with no lower bounds on arrival times and no waiting decisions, yielding a spectrum of temporal restrictions from tightly constrained time windows to time-dependent routing without time windows. Despite substantial progress, the weakly constrained end of this spectrum remains challenging for exact methods: to date, several benchmark instances of the \TDTSP{} with 25 or more customers remain open.

In this paper, we study the \TDTSPTW{} and aim to extend the range of weakly constrained instances that can be solved exactly.

\subsection{Motivation}

The method of \cite{Lera-RomeroMirandaBrontSoulignac2022}, originally proposed for the \TDTSPTW{} with duration minimization, relies on dynamic programming with ng-relaxations and builds on earlier approaches for related TSPTW variants \citep[see][]{BaldacciMingozziRoberti2012,TilkIrnich2017}. It follows a multi-phase framework based on a common optimization problem: a $\Profit$-\TDTSPTW{} in which visiting vertex $\Vertex$ yields a benefit $\Profit_\Vertex$, and tours are evaluated through a reduced-cost criterion combining travel times and vertex benefits. Its phases solve progressively stronger relaxations characterized by a vertex-based ng-memory structure defining path feasibility.

The first phase computes $\Profit$ as the optimal dual solution of the final restricted master problem obtained through column generation. Each pricing problem is a $\Profit$-\TDTSPTW{} with a small ng-memory structure, solved by forward labeling to identify negative reduced-cost tours. The second phase iteratively enlarges the ng-memories to forbid cycles detected in minimum reduced-cost tours. The dual vector $\Profit$ is kept fixed throughout this phase, and the resulting sequence of strengthened $\Profit$-\TDTSPTW{}s is solved through bidirectional labeling. Since stronger ng-memory structures reduce dominance opportunities, these pricing problems become progressively harder to solve despite using the same labeling machinery. Finally, the third phase computes an optimal tour through an exact forward labeling algorithm within a bounding framework. It first generates feasible backward paths by solving a relaxed $\Profit$-\TDTSPTW{} using $\Profit$ and the final ng-memory structure, and then performs a best-first forward search guided by compatible backward completions.

While improving upon previous solvers for the \TDTSPTW{}, this framework exhibits several limitations that become particularly relevant for instances with loose time windows. First, the pricing problems required to compute $\Profit$ in the initial phase are computationally demanding even with relatively small ng-memory structures. Since column generation is performed under a fixed time budget (one third of the total computation time), $\Profit$ may be obtained before convergence, potentially limiting its effectiveness in guiding subsequent phases. Moreover, reusing $\Profit$ across phases restricts the ability of the method to adapt the reduced-cost information to stronger ng-memory structures, since subsequent decisions remain based on the initial relaxation. Finally, iterative ng-memory enlargement may require substantial computational effort without proportional gains in pruning efficiency. This limitation becomes more pronounced for loose time windows, where many alternative cycles with comparable costs may remain available after individual cycles are forbidden. The sequential organization of the framework, with predefined phases and fixed time budgets, further limits the integration of search strategies such as branch-and-bound.

\subsection{Our contributions}

This paper develops an exact solution framework for the \TDTSPTW{} that extends the range of instances solvable to optimality under loose time windows while remaining effective across different levels of time-window tightness. Motivated by the limitations of the method of \cite{Lera-RomeroMirandaBrontSoulignac2022}, we redesign the interaction among column generation, ng-memory augmentation, and exact search, allowing information generated by each component to guide the others instead of treating them as independent sequential stages.

Rather than applying these procedures once, we embed them into an iterative framework. In each iteration, a dual vector $\Profit$ is obtained through column generation using the current ng-memory structure. The ng-memories are then augmented to forbid cycles appearing in tours obtained by solving progressively stronger $\Profit$-\TDTSPTW{}s. Finally, an exact search, guided by completion bounds derived from $\Profit$ and the current ng-memory structure, is executed for a limited time budget. If optimality is not proven, the process continues with the strengthened structure, enabling $\Profit$ and ng-memories to co-evolve.

As in the framework of \cite{Lera-RomeroMirandaBrontSoulignac2022}, a key challenge is solving the $\Profit$-\TDTSPTW{} instances in the initial column generation phase. Under loose time windows, labeling methods explore a large state space even with small ng-memories. To address this issue, we embed a completion-bound-driven sparsification mechanism within column generation. After solving a $\Profit$-\TDTSPTW{} instance via forward labeling, a backward labeling may temporarily remove arcs with poor completion bounds based on $\Profit$. These reductions decrease the effective network size and accelerate subsequent pricing problems. The mechanism is used both in heuristic pricing, to quickly generate a rich set of columns, and in exact pricing, where it provides an exact variable-fixing procedure that eliminates arcs that cannot belong to an optimal tour. Additionally, the incumbent solution is continuously improved through a primal heuristic, producing tighter completion bounds that enhance sparsification and variable fixing.

The proposed framework further exploits time dependency through an arc-position ng-memory structure, where the additional ng-memory required to forbid a cycle depends on its arcs and position within a tour. This reflects the fact that the cost of traversing a cycle varies with departure time, with position serving as a proxy. Consequently, ng-memory structures can be refined more selectively, allowing more cycles to be processed before reaching a prohibitive combinatorial explosion.

Extensive computational experiments demonstrate that our approach significantly extends the capabilities of existing exact methods. It solves all benchmark instances considered in \cite{Lera-RomeroMirandaBrontSoulignac2022} and \cite{Fontaine2024} with up to 45 customers, whereas the method of \cite{Lera-RomeroMirandaBrontSoulignac2022} fails to solve some instances without time windows with only 20 customers. These results establish a new state of the art for the exact solution of the \TDTSPTW{} in regimes with loose or absent time-window constraints.

\section{Problem Formulation and Solution Framework}
\label{sec:problem statement}

This section introduces the notation and optimization problems underlying the proposed framework. We first define the \TDTSPTW{} and its ng-relaxation, then present the master problem ($\Master$) used to obtain dual information and certify optimality. Finally, we introduce the $\Profit$-\TDTSPTW{}, the pricing problem solved throughout the framework, and conclude with an overview of the algorithm.

\subsection{Problem Statement}

Throughout this article, we write $[j]=[0,j]$, $\Range{i,j}=[i,j]\cap\mathbb{N}$, and $\Range{j}=\Range{0,j}$ for $i,j\in\mathbb{R}$. Consider a \emph{transport network} $\Network$ represented by a complete digraph with vertex set $\Range{\CustCount+1}$, where vertices $0$ and $\CustCount+1$ denote the \emph{start} and \emph{end depots}, and vertices in $\Range{1,\CustCount}$ denote \emph{customers}. A \emph{(forward) path} $\Route$ is a nonempty vertex sequence $\Route=\Sequence{\Vertex_0,\ldots,\Vertex_k}$ with $\Vertex_0=0$. A path is \emph{elementary} if it contains no repeated vertices, and a \emph{tour} if $k=\Vertex_k=\CustCount+1$; thus, an elementary tour visits each customer exactly once.

A vehicle operates in $\Network$ within a \emph{planning horizon} $\Horizon\in\mathbb{R}$. Each arc $\Vertex\Wertex$ has a piecewise-linear \emph{travel time} function $\TravelTime(\Vertex\Wertex,\Time)$ with domain $[\Horizon]$, representing the traversal time when departing from $\Vertex$ at time $\Time$. These functions satisfy the \emph{FIFO property}, namely, $\Time' + \TravelTime(\Vertex\Wertex,\Time+\Time') \geq \TravelTime(\Vertex\Wertex,\Time)$ for $\Time'\in [\Horizon-\Time]$.

\begin{remark}\label{rem:time-windows}
Under the adopted definition of $\TravelTime$, time windows can be incorporated into travel time functions. Specifically, for a \TDTSPTW{} instance with window $[\Release(\Vertex),\Deadline(\Vertex)]$ at customer $\Vertex$, each incoming arc $\Vertex\Wertex$ can be modified to delay arrivals before $\Release(\Wertex)$ by setting $\TravelTime(\Vertex\Wertex,\Time)=\Release(\Wertex)-\Time$ whenever $\Time+\TravelTime(\Vertex\Wertex,\Time)<\Release(\Wertex)$, and to make arrivals after $\Deadline(\Wertex)$ infeasible by setting $\TravelTime(\Vertex\Wertex,\Time)=\Horizon-\Time+1$ whenever $\Time+\TravelTime(\Vertex\Wertex,\Time)>\Deadline(\Wertex)$. Thus, any \TDTSPTW{} instance can be viewed as a \TDTSP{} instance, and conversely by assigning $[\Horizon]$ as the time window of every customer. Nevertheless, we explicitly model time windows in the arc-position states below, as they provide a compact representation of temporal restrictions and a natural characterization of benchmark instances according to window tightness.
\end{remark}

In the proposed framework, time windows and ng-memories are associated with arc-position states rather than vertices. Thus, for $\Vertex,\Wertex\in\Range{\CustCount+1}$ and $\Position \in \Range{\CustCount}$, each pair $(\Vertex\Wertex,\Position)$ is associated with a (possibly empty) \emph{time window} $[\Release(\Vertex\Wertex,\Position),\Deadline(\Vertex\Wertex,\Position)]\subseteq[\Horizon]$ and an \emph{ng-memory} $\Memory(\Vertex\Wertex,\Position)\subseteq\Range{\CustCount+1}$. The time window imposes that traversing $\Vertex\Wertex$ at position $\Position$ starts no earlier than $\Release(\Vertex\Wertex,\Position)$ and is completed no later than $\Deadline(\Vertex\Wertex,\Position)$. The ng-memory requires every cycle in a path $\Route = \Sequence{\Vertex_0, \ldots, \Vertex_k}$ to contain an arc-position state whose memory excludes the repeated vertex. Formally, $\Route$ is \emph{$\Memory$-feasible} if, for every $i,j\in\Range{k}$ with $i<j$ and $\Vertex_i=\Vertex_j$, there exists $\Position\in\Range{i,j-1}$ such that $\Vertex_i\notin\Memory(\Vertex_\Position\Vertex_{\Position+1},\Position)$. Note that every elementary tour is $\Memory$-feasible.

For $\Position \in \Range{k}$, the \emph{earliest arrival time} at $\Vertex_\Position$ along a path $\Route=\Sequence{\Vertex_0,\ldots,\Vertex_k}$ is recursively defined as
\begin{equation}
 \ArrivalTime(\Route, \Position) = \begin{cases}
         0 & \text{if } \Position = 0, \\
         \ArrivalTime(\Route, \Position-1) + \TravelTime(\Vertex_{\Position-1}\Vertex_{\Position}, \max\{\ArrivalTime(\Route, \Position-1), \Release(\Vertex_{\Position-1}\Vertex_\Position, \Position-1)\}) & \text{otherwise,}
        \end{cases}\label{eq:earliest-arrival-time}
\end{equation}
where the maximum term accounts for waiting times. Path $\Route$ is \emph{ng-feasible} if it is $\Memory$-feasible and satisfies $\ArrivalTime(\Route,\Position)\leq\Deadline(\Vertex_{\Position-1}\Vertex_\Position,\Position-1)$ for all $\Position\in\Range{1,k}$. We define $\ArrivalTime(\Route)=\ArrivalTime(\Route,k)$ if $\Route$ is ng-feasible and $\ArrivalTime(\Route)=\Horizon+1$ otherwise, referring to $\ArrivalTime(\Route)$ as the \emph{makespan} of $\Route$.

The \TDTSPTW{} is the problem of finding an ng-feasible elementary tour of minimum makespan in transport network $\Network$. The ng-memory structure is not part of the optimization problem itself, but defines ng-relaxations of the elementary tour constraint exploited throughout our framework.

\subsection{Certification of Optimality}

Suppose that an elementary tour $\Route^*$ is available. Our framework repeatedly seeks to certify that $\Route^*$ is optimal by considering the set $\RouteUniverse(\Network,\Route^*)$ of ng-feasible tours in $\Network$ with makespan strictly smaller than $\ArrivalTime(\Route^*)$. If no affine combination of tours in this set covers all customers, then no improving elementary tour exists, and $\Route^*$ is optimal. This motivates the following linear program, where $\Repetitions_{\Vertex\Route}$ denotes the number of visits to vertex $\Vertex$ by tour $\Route\in\RouteUniverse(\Network,\Route^*)$, and $\DecisionVar_\Route$ denotes its coefficient in the affine combination.
\begin{align}
    & & \Master(\Network,\Route^*)\colon \min &  \sum_{\Route \in \RouteUniverse(\Network, \Route^*)} \ArrivalTime(\Route) \DecisionVar_\Route \label{eq:obj}\\
    & & \textrm{s.t.} & \sum_{\Route \in \RouteUniverse(\Network, \Route^*)} \DecisionVar_\Route \leq 1 \label{eq:unique}\\
    & &  & \sum_{\Route \in \RouteUniverse(\Network, \Route^*)} \Repetitions_{\Vertex\Route} \DecisionVar_\Route \geq 1 & & \text{for every } \Vertex \in \Range{1,\CustCount}\label{eq:degree}\\
    & &     & \DecisionVar_\Route \in [0, 1]  & & \text{for every }  \Route \in \RouteUniverse(\Network, \Route^*).\label{eq:domain}
\end{align}
The objective function~\eqref{eq:obj} minimizes the weighted makespan of an affine combination of tours in $\RouteUniverse(\Network,\Route^*)$. Constraints~\eqref{eq:unique} and~\eqref{eq:domain} define the affine combination, while constraints~\eqref{eq:degree} ensure that every customer is covered at least once.

Unlike classical set-partitioning formulations, $\RouteUniverse(\Network,\Route^*)$ depends on the incumbent solution and includes only tours improving $\Route^*$. Therefore, $\Master(\Network,\Route^*)$ may be infeasible. In this case, no improving elementary tour exists, and $\Route^*$ is optimal for the \TDTSPTW{}. Initially, $\Route^*$ is an artificial tour with makespan $\Horizon+1$. Thus, if no improving tour exists, infeasibility of $\Master(\Network,\Route^*)$ certifies that the \TDTSPTW{} on $\Network$ is infeasible. Herafter, we omit notation arguments that are clear from context; for example, $\Network$ and $\Route^*$ are omitted from $\RouteUniverse$ and $\Master$ whenever unambiguous.

\subsection{The \texorpdfstring{$\Profit$}{dual}-\texorpdfstring{\TDTSPTW}{TDTSPTW}}
\label{sec:profit weighted tdtsptw}

All major components of the proposed framework rely on the \emph{$\Profit$-\TDTSPTW{}}, parameterized by a vector of vertex benefits $\Profit\in\mathbb{R}^{\CustCount+1}$, typically obtained from the dual variables of $\Master$. Given $\Network$, $\Route^*$, and $\Profit$, the problem consists of finding a tour $\Route\in\RouteUniverse$ minimizing
\begin{displaymath}
 \RedArrivalTime(\Network, \Profit, \Route) = \ArrivalTime(\Route) - \sum_{\Vertex \in \Range{\CustCount+1}}\Repetitions_{\Vertex\Route}\Profit_\Vertex.
\end{displaymath}
When $\Profit$ is dual feasible for a restricted $\Master$, $\RedArrivalTime$ equals the \emph{reduced cost} of $\Route$. Hence, we refer to $\Profit$ as a \emph{dual vector} and to $\RedArrivalTime$ as the \emph{reduced cost}, even if $\Profit$ is not dual feasible.

Each dual vector $\Profit$ provides a lower bound on the makespan of any elementary tour $\OtherRoute\in\RouteUniverse$, since
\begin{displaymath}
 \PricingLowerBound(\Network, \Route^*, \Profit) = \min\{\RedArrivalTime(\Route) \mid \Route \in \RouteUniverse\} + \TotalProfit \leq \RedArrivalTime(\OtherRoute) + \TotalProfit = \ArrivalTime(\OtherRoute), \label{eq:pricing-lower-bound}
\end{displaymath}
where $\TotalProfit=\sum_{\Vertex\in\Range{\CustCount+1}}\Profit_\Vertex$. If $\PricingLowerBound\geq\ArrivalTime(\Route^*)$, then $\Route^*$ is optimal. We refer to $\PricingLowerBound$ as the \emph{$\Profit$-bound} of $(\Network,\Route^*)$.

\subsection{Overview of the framework}

Algorithm~\ref{alg:solver} provides an overview of the proposed framework. Starting from an initial incumbent tour, each iteration attempts to certify its optimality by strengthening $\Master$ through four complementary mechanisms: increasing the $\Profit$-bound, improving the incumbent solution, reducing the transport network, and augmenting the ng-memory structure. These mechanisms interact with the exact search procedure and reinforce one another until the incumbent tour is certified optimal.

\begin{algorithm}[tbh]
\caption{Solver (high-level description)}\label{alg:solver}
\begin{algorithmic}[1]
    \Require a complete transport network $\Network$

    \Ensure an optimal elementary tour $\Route^*$ in $\Network$, or an artificial one if $\Network$ has no elementary tours

    \State \Let $\LowerBound = 0$, and \Let $\Route^*$ be an artificial elementary tour with $\ArrivalTime(\Route^*) = \Horizon+1$\label{alg:solver:init}

    \While{$\LowerBound < \ArrivalTime(\Route^*)$}\label{alg:solver:loop start}

        \State \textbf{Increase the $\Profit$-bound:} invoke the CG method (Section~\ref{sec:column generation}) to obtain a solution of $\Master$ with support $\RouteSet$, whose corresponding dual solution $\Profit$ yields a strong $\Profit$-bound $\PricingLowerBound$\label{alg:solver:CG}

        \State \textbf{Improve incumbent:} execute the primal heuristic (Appendix~\ref{app:primal}) on selected columns of the restricted $\Master$ obtained in Step~\ref{alg:solver:CG} to update $\Route^*$.\label{alg:solver:primal}

        \State \textbf{Reduce the network:} run the variable fixing method (Section~\ref{sec:variable fixing}) to remove from $\Network$ the arc-position combinations that cannot belong to any elementary tour $\Route$ satisfying $\ArrivalTime(\Route) < \ArrivalTime(\Route^*)$.\label{alg:solver:VF}

        \State \textbf{Augment ng-memories:} update ng-memories to first forbid cycles on each tour of $\RouteSet$, and then on the best tours obtained from repeated solutions of the $\Profit$-\TDTSPTW{} (Section~\ref{sec:memory augmentation})\label{alg:solver:DNA}

        \State \textbf{Exact search:} try to find the optimal solution $\Route^*$ by performing an exact search over the elementary tours in $\Network$, taking advantage of the paths of Step~\ref{alg:solver:DNA} to compute completion bounds.\label{alg:solver:enumerate}\label{alg:solver:loop end}

    \EndWhile
\end{algorithmic}
\end{algorithm}

A stronger dual vector $\Profit$ provides tighter completion bounds, benefiting all procedures that rely on these bounds, including sparsification, variable fixing, exact search, and ng-memory augmentation. In particular, greater $\Profit$-bounds enable variable fixing to forbid additional arc-position combinations from the network, reducing the search space and simplifying subsequent $\Profit$-\TDTSPTW{} instances. Conversely, greater $\Profit$-bounds are obtained by removing tours from $\Master$ via ng-memory augmentation, enlarging the search space of subsequent $\Profit$-\TDTSPTW{} instances. Algorithm~\ref{alg:solver} balances these effects, alternating between the four mechanisms.

The framework does not apply variable fixing during early iterations of the initial column generation process because the exact completion bounds provide limited pruning at this stage. Consequently, variable fixing would require substantial computational effort to remove only a few arc-position combinations from the network. Even with heuristic labeling relying on strengthened dominance rules, solving pricing problems on the complete network is computationally demanding. To mitigate this issue, we embed a sparsification mechanism that applies the completion-bound principle underlying variable fixing within a heuristic scheme. Heuristic completion bounds derived from the non-dominated paths of the previous pricing iteration are used to remove non-promising arc-position combinations from the network. Sparsification may discard arc-position combinations that belong to every optimal tour; in such cases, the full transport network is eventually restored and sparsification is reapplied if appropriate. Overall, fewer pricing problems are solved on the complete network.

Early-stage sparsification is complemented by the primal heuristic to tighten completion bounds. Their combination accelerates the transition from the initial exploratory phase to a state where variable fixing becomes effective, making the first iteration of Algorithm~\ref{alg:solver} computationally viable.

\section{Solving a \texorpdfstring{$\Profit$}{dual}-\texorpdfstring{\TDTSPTW{}}{TDTSPTW}}
\label{sec:pricing}

To solve $\Profit$-\TDTSPTW{}s, we implement a labeling algorithm that implicitly explores a tree of paths $\Tree(\Network,\Route^*)$. Since the algorithm follows a standard labeling framework, we describe only the components specific to our implementation and required for reproducibility.

The root of $\Tree$ is the path $\Sequence{0}$ containing only the start depot, and its leaves correspond to the tours in $\RouteUniverse$. A node at depth $\Position\leq\CustCount$ represents a path $\Route=\Sequence{\Vertex_0,\ldots,\Vertex_\Position}$, whose children are the ng-feasible extensions $\Route+\Sequence{\Vertex}$ for $\Vertex\in\Range{\CustCount+1}$. Each node is represented by a \emph{forward label} $\Label$ requiring $O(1)$ words of memory. The label stores a pointer $\Prev(\Label)$ to its parent, the \emph{position} $\Position(\Label) = \Position$, the \emph{last vertex} $\Vertex(\Label) = \Vertex_\Position$, the \emph{makespan} $\ArrivalTime(\Label)=\ArrivalTime(\Route)$, the \emph{profit} $\Profit(\Label)=\sum_{i=0}^{\Position}\Profit_{\Vertex_i}$, and the \emph{memory} $\Memory(\Label)=\{\Vertex\in\Range{\CustCount+1}\mid \Route+\Sequence{\Vertex}\text{ is not ng-feasible}\}$, implemented as a bitset. Unless otherwise stated, nodes, labels, and the paths they represent are used interchangeably.

For $\Wertex \in \Range{\CustCount+1}$, the path $\Route+\Sequence{\Wertex}$ is a child of $\Route$ if $\Wertex\notin\Memory(\Label)$, $\ArrivalTime(\Route+\Sequence{\Wertex})\leq\Deadline(\Vertex_\Position\Wertex,\Position)$, and $\ArrivalTime(\Route+\Sequence{\Wertex})<\ArrivalTime(\Route^*)$. The corresponding label $\Label'$ is obtained from $\Label$ by setting $\Prev(\Label')=\Label$, $\Position(\Label')=\Position+1$, $\Vertex(\Label')=\Wertex$, $\Profit(\Label')=\Profit(\Label)+\Profit_{\Wertex}$, $\ArrivalTime(\Label') = \ArrivalTime(\Label) + \TravelTime(\Vertex_\Position\Wertex,\max\{\ArrivalTime(\Label),\Release(\Vertex_\Position\Wertex,\Position)\})$, and $\Memory(\Label') = (\Memory(\Label)\cup\{\Vertex_\Position\})\cap\Memory(\Vertex_\Position\Wertex,\Position)$. Travel times are evaluated in $O(1)$ time using an arc-specific lookup table that identifies the relevant piecewise-linear segments of each travel-time function. Consequently, each label extension can be performed in $O(1)$ time.

Dominance rules are used to prune nodes without affecting optimality. For paths $\Route$ and $\Route'$ ending at the same position, $\Route$ \emph{dominates} $\Route'$ if, for every $\Route'+\OtherRoute\in\RouteUniverse$, we have $\Route+\OtherRoute\in\RouteUniverse$ and $\RedArrivalTime(\Route+\OtherRoute)\leq\RedArrivalTime(\Route'+\OtherRoute)$. We apply the following rule to efficiently identify dominated labels.

\begin{domrule}\label{rule:dominance}
 If $\Position(\Label) = \Position(\Label')$, $\Vertex(\Label) = \Vertex(\Label')$, $\ArrivalTime(\Label) \leq \ArrivalTime(\Label')$, $\Profit(\Label) \geq \Profit(\Label')$, and $\Memory(\Label) \subseteq \Memory(\Label')$, then $\Label$ dominates $\Label'$.
\end{domrule}

The \emph{(forward) labeling algorithm} explores $\Tree$ using breadth-first search. Dominance is not applied to labels reaching the end depot, allowing the generation of additional tours. These tours enrich the column set of $\Master$, accelerating column generation convergence, and are used to enlarge the ng-memories.

\paragraph{Heuristic labeling.}
We implement two heuristic variants of the labeling algorithm to accelerate early column generation. The \emph{relax-all} variant removes the arrival-time and memory-inclusion conditions from Rule~\ref{rule:dominance}, retaining only profit comparison, whereas \emph{relax-ng} removes only the memory-inclusion condition. For convenience, we use \eqref{eq:pricing-lower-bound}, retaining the notation $\PricingLowerBound$, to estimate a lower bound for the \TDTSPTW{}. This estimate is not guaranteed to be a valid lower bound.

Relax-ng may remain computationally demanding because its dominance rule still depends on a trade-off between arrival time and profit. The continuous nature of $\Profit$ makes this relation sensitive to small variations in dual values. This effect occurs for all time-window widths but is amplified for loose time windows, where fewer labels are eliminated by temporal infeasibility.

\subsection{Backward labeling}
\label{sec:backward labeling}

In this section, we present a backward labeling algorithm analogous to the forward procedure, describing only the differences. Since solving the \TDTSPTW{} by backward labeling requires an additional state variable to account for the possible arrival times at the end depot \citep{Lera-RomeroMirandaBrontSoulignac2022}, we instead fix the target arrival time to the incumbent makespan $\ArrivalTime(\Route^*)$. This avoids the larger state space while providing all the information required for variable fixing, sparsification, and ng-memory augmentation.

For each arc $\Vertex\Wertex$, let the \emph{backward travel-time} function $\BackTravelTime(\Vertex\Wertex,\Time)$ denote the travel time corresponding to the latest departure that reaches $\Wertex$ by time $\Time\in[\Horizon]$. Formally, $\BackTravelTime(\Vertex\Wertex, \Time) = \Time - \max\{\Time' \in [\Horizon] \mid \Time' + \TravelTime(\Vertex\Wertex, \Time') \leq \Time\}$. A \emph{backward path} is a nonempty suffix $\Route=\Sequence{\Vertex_{\CustCount+1-k},\ldots,\Vertex_{\CustCount+1}}$ of a tour. Elementarity and $\Memory$-feasibility are defined analogously to the forward setting, and we omit the qualifier \emph{backward} whenever the context is clear. Let $\DepartureTime(\Route,\Position)$ denote the \emph{latest departure time} from $\Vertex_\Position$ that allows the remaining suffix of $\Route$ to reach the end depot no later than $\ArrivalTime(\Route^*)$. It is computed recursively as follows:
\begin{equation}
 \DepartureTime(\Route, \Position) = \begin{cases}
         \ArrivalTime(\Route^*) & \text{if } \Position = \CustCount + 1, \\
         \DepartureTime(\Route, \Position+1) -  \BackTravelTime(\Vertex_{\Position}\Vertex_{\Position + 1}, \min\{\DepartureTime(\Route, \Position+1), \Deadline(\Vertex_{\Position}\Vertex_{\Position+1}, \Position)\}) & \text{otherwise.}
        \end{cases} \label{eq:latest-departure-time}
\end{equation}
A path $\Route$ is \emph{bng-feasible} if it is $\Memory$-feasible and satisfies $\DepartureTime(\Route,\Position)\geq\Release(\Vertex_{\Position}\Vertex_{\Position+1},\Position)$ for every $\Position\in\Range{\CustCount+1-k,\CustCount}$. We define the \emph{departure time} of $\Route$ as $\DepartureTime(\Route)=\DepartureTime(\Route,k)$ if $\Route$ is bng-feasible, and $\DepartureTime(\Route)=-1$ otherwise. The set $\BackRouteUniverse(\Route^*)$ of bng-feasible tours consists of the tours in $\RouteUniverse$ and those ng-feasible with makespan equal to $\ArrivalTime(\Route^*)$.

Let $\BackTree(\Network,\Route^*)$ denote the tree rooted at $\Sequence{\CustCount+1}$, whose leaves correspond to the tours in $\BackRouteUniverse$. A node at depth $\Position\leq\CustCount$ represents a backward path $\Route=\Sequence{\Vertex_{\CustCount+1-\Position},\ldots,\Vertex_{\CustCount+1}}$, whose children are the bng-feasible extensions $\Sequence{\Vertex}+\Route$ for $\Vertex\in\Range{\CustCount+1}$. Each node is represented by a \emph{backward label} $\Label$ storing a pointer $\Next(\Label)$ to its parent, the \emph{position} $\Position(\Label)=\CustCount+1-\Position$, the \emph{initial vertex} $\Vertex(\Label) = \Vertex_{\CustCount+1-\Position}$, the \emph{departure time} $\DepartureTime(\Label)=\DepartureTime(\Route)$, the \emph{profit} $\Profit(\Label)=\sum_{i=0}^{\Position}\Profit_{\Vertex_{\CustCount+1-i}}$, and the \emph{memory} $\Memory(\Label)= \{\Vertex\in\Range{\CustCount+1}\mid \Sequence{\Vertex}+\Route \text{ is not bng-feasible}\}$. Unless otherwise stated, backward nodes, labels, and the paths they represent are used interchangeably.

Recall that $\Repetitions_{\Vertex\Route}$ denotes the number of visits of vertex $\Vertex$ in tour $\Route$.  Let $\RedDepartureTime(\Network, \Profit, \Route) = \ArrivalTime(\Route^*) - \DepartureTime(\Route) - \sum_{\Vertex \in \Range{\CustCount+1}} \Repetitions_{\Vertex\Route}\Profit_\Vertex$ be the \emph{backward reduced cost} of $\Route$.  For backward paths $\Route$ and $\Route'$ at the same position, $\Route$ \emph{dominates} $\Route'$ if, for every $\OtherRoute+\Route'\in\BackRouteUniverse$, we have $\OtherRoute+\Route\in\BackRouteUniverse$ and $\RedDepartureTime(\OtherRoute+\Route) \leq \RedDepartureTime(\OtherRoute+\Route')$

\begin{domrule}\label{rule:back-dominance}
 If $\Position(\Label) = \Position(\Label')$, $\Vertex(\Label) = \Vertex(\Label')$, $\DepartureTime(\Label) \geq \DepartureTime(\Label')$, $\Profit(\Label) \geq \Profit(\Label')$, and $\Memory(\Label) \subseteq \Memory(\Label')$, then $\Label$ dominates $\Label'$.
\end{domrule}

The \emph{backward labeling algorithm} explores $\BackTree$ without applying dominance to complete tours, allowing the generation of additional tours for ng-memory augmentation. It can also be interpreted as solving a \emph{backward $\Profit$-\TDTSPTW{}} that finds a tour $\Route \in \BackRouteUniverse$ minimizing $\RedDepartureTime$.  However, unlike the tour in $\RouteUniverse$ minimizing $\RedArrivalTime$, $\Route$ cannot be used to derive a lower bound for the \TDTSPTW{} analogous to $\PricingLowerBound$ \eqref{eq:pricing-lower-bound}. The heuristic variants relax-all and relax-ng are defined analogously to their forward counterparts.

\subsection{Bounded labeling}
\label{sec:bounded labeling}

Bounded labeling is an acceleration technique for solving sequences of $\Profit$-\TDTSPTW{} instances over transport networks with progressively smaller $\RouteUniverse$ sets while keeping $\Profit$ fixed \citep{BaldacciMingozziRoberti2012,TilkIrnich2017,Lera-RomeroMirandaBrontSoulignac2022}. Besides dominance, it prunes labels that cannot be extended into tours satisfying a completion criterion. We adapt it below to the forward and backward labeling algorithms.

Let $\Network$ and $\Network'$ be transport networks whose $\RouteUniverse$ sets are comparable by inclusion, and denote by $\Network^-$ and $\Network^+$ the networks inducing the smaller and larger sets, respectively. A forward label $\Label \in \Tree(\Network)$ and a backward label $\Mabel \in \BackTree(\Network')$ are \emph{compatible} if $\Position(\Label)=\Position(\Mabel)$, $\Vertex(\Label)=\Vertex(\Mabel)$, $\Memory(\Label)\cap\Memory(\Mabel)=\emptyset$, and $\ArrivalTime(\Label)\leq\DepartureTime(\Mabel)$. The first three conditions ensure that $\Route=\Label\oplus\Mabel$ is a bng-feasible tour of $\Network^+$, where $\oplus$ concatenates the paths represented by the two labels by merging their common vertex. The last condition guarantees $\DepartureTime(\Route)\geq0$ and $\ArrivalTime(\Route)\leq\ArrivalTime(\Route^*)$, implying that $\Route \in \BackRouteUniverse(\Network^+)$. Moreover, $\ArrivalTime(\Route)=\ArrivalTime(\Route^*)$ can only occur when $\Position(\Label)<\CustCount+1$.  The corresponding \emph{completion value} is
\begin{align}
\RedCost(\Label,\Mabel) &= \left[\ArrivalTime(\Label) - \Profit(\Label)\right]  + \left[\ArrivalTime(\Route^*) - \DepartureTime(\Mabel) - \Profit(\Mabel)\right] + \Profit_{\Vertex(\Label)} - \RedArrivalTime(\Route^*) \label{eq:completion:value} \\
  &\leq \ArrivalTime(\Label \oplus \Mabel) - \sum_{\Vertex \in \Range{\CustCount+1}} \Repetitions_{\Vertex(\Label\oplus\Mabel)}\Profit_{\Vertex} - \RedArrivalTime(\Route^*) = \RedArrivalTime(\Label \oplus \Mabel) - \RedArrivalTime(\Route^*).\notag
\end{align}

\begin{observation}
 if $\RedCost(\Label, \Mabel) > 0$, then $\RedArrivalTime(\Label\oplus\Mabel) > \RedArrivalTime(\Route^*)$.
\end{observation}

If $\Route^*$ is not optimal in $\Network^-$, there exists an elementary tour $\Route\in\RouteUniverse(\Network^-)$ with $\RedArrivalTime(\Route)<\RedArrivalTime(\Route^*)$ and $\RedDepartureTime(\Route)\leq\RedDepartureTime(\Route^*)$. Moreover, for every $\Position\in\Range{\CustCount+1}$, $\Route$ can be decomposed as $\Route=\Label\oplus\Mabel$ into compatible labels $\Label\in\Tree(\Network)$ and $\Mabel\in\BackTree(\Network')$ satisfying $\Position(\Label)=\Position(\Mabel)=\Position$ and $\RedCost(\Label,\Mabel)\leq0$. Although these labels may be removed by dominance, Rules~\ref{rule:dominance} and~\ref{rule:back-dominance} imply the following result.

\begin{proposition}\label{prop:completion bound}
Let $\Position \in \Range{\CustCount+1}$. If $\Route^*$ is not optimal in $\Network^-$, then there exist non-dominated compatible labels $\Label \in \Tree(\Network)$ and $\Mabel \in \BackTree(\Network')$ satisfying $\Position(\Label) = \Position(\Mabel) = \Position$ and $\RedCost(\Label,\Mabel) \leq 0$.
\end{proposition}

The \emph{completion bound} of a forward label $\Label$ induced by a set of backward labels $\BackLabelSet$ is $\CompletionBound(\Label) = \min\{\RedCost(\Label,\Mabel) \mid \Mabel \in \BackLabelSet\}$.  Given $\BackLabelSet$, the \emph{forward bounded labeling algorithm} extends the labeling procedure by discarding the labels with $\CompletionBound > 0$. By Proposition~\ref{prop:completion bound}, when executed on $\Network^-$ with $\BackLabelSet$ containing all non-dominated labels of $\BackTree(\Network^+)$, the algorithm either finds a tour in $\RouteUniverse(\Network^-)$ with $\RedArrivalTime < \RedArrivalTime(\Route^*)$ or certifies the optimality of $\Route^*$.  Since labels with $\CompletionBound >0$ are discarded, the search is restricted to a subset of the tours computed by the labeling method.  Therefore, the resulting lower bound $\RedArrivalTime(\Route)+\TotalProfit$, where $\Route$ minimizes $\RedArrivalTime$ among the explored tours, is never weaker than $\PricingLowerBound$ in \eqref{eq:pricing-lower-bound}. For simplicity, we use the same notation $\PricingLowerBound$ for this bound, as the meaning follows from the context.

Completion bounds of backward labels and the backward bounded labeling algorithm are defined analogously.  The latter obtains a tour $\Route \in \BackRouteUniverse(\Network^-)$ minimizing $\RedDepartureTime(\Route)$.  Unlike the forward case, however, $\RedDepartureTime(\Route)$ does not provide a lower bound for the \TDTSPTW{} analogous to $\PricingLowerBound$, as discussed in Section~\ref{sec:backward labeling}.

An immediate consequence of Proposition~\ref{prop:completion bound} is that bounded labeling can be applied iteratively over transport networks with progressively smaller $\RouteUniverse$ sets. Specifically, let $\Network_0,\ldots,\Network_k$ be transport networks satisfying $\RouteUniverse(\Network_{i+1})\subseteq\RouteUniverse(\Network_i)$ for $i\in\Range{k-1}$. Then, the non-dominated labels generated by a bounded labeling algorithm on $\Network_i$ can be used to induce completion bounds when solving $\Network_{i+1}$. This requires that the initial label set for $\Network_0$ contains all corresponding non-dominated labels.

The efficiency of bounded labeling depends on the incumbent tour $\Route^*$, the dual vector $\Profit$, and the ng-memory structure. When the gap between $\PricingLowerBound(\Network,\Route^*,\Profit)$ and $\ArrivalTime(\Route^*)$ is large, relatively few completion bounds are positive, so completion tests provide limited pruning while introducing computational overhead. As this gap narrows, completion bounds increase, allowing bounded labeling to discard a larger fraction of labels. The primal heuristic primarily improves $\ArrivalTime(\Route^*)$, whereas column generation and ng-memory enlargement primarily improve $\PricingLowerBound$.

\subsubsection{Variable fixing and sparsification}
\label{sec:variable fixing}

Completion bounds are also used in Step~\ref{alg:solver:VF} of Algorithm~\ref{alg:solver} to forbid arc-position states that do not belong to any elementary tour improving $\Route^*$, following the variable fixing framework of \citet{IrnichDesaulniersDesrosiersHadjarIJoC2010}. The same decomposition argument used in the previous section yields the following proposition.

\begin{proposition}
\label{prop:variable fixing}
Let $\Position \in \Range{\CustCount}$ and $\Vertex,\Wertex\in\Range{\CustCount+1}$. If an elementary tour $\Route\in\RouteUniverse$ traverses arc $\Vertex\Wertex$ at position $\Position$, then there exist compatible labels $\Label \in \Tree$ and $\Mabel \in \BackTree$ with $\Position(\Label)=\Position(\Mabel)=\Position$ such that $\Label$ and $\Next(\Mabel)$ are non-dominated, and $\DepartureTime(\Next(\Mabel)) \geq\DepartureTime(\Route,\Position)$.
\end{proposition}

Given a set $\LabelSet$ of forward labels, the \emph{variable fixing algorithm} extends the backward bounded labeling algorithm by computing, for each arc-position state $(\Vertex\Wertex,\Position)$, the maximum value $\Time$ of $\DepartureTime(\Next(\Mabel))$ among the enumerated backward labels satisfying $\Vertex(\Mabel) = \Vertex$, $\Vertex(\Next(\Mabel)) = \Wertex$, $\Position(\Mabel) = \Position$, and $\CompletionBound(\Mabel)\leq0$. The deadline $\Deadline(\Vertex\Wertex,\Position)$ is updated to $\Time$, or set to $-1$ otherwise, thereby declaring the state infeasible. By Proposition~\ref{prop:variable fixing}, every elementary tour in $\RouteUniverse(\Network)$ remains feasible and preserves its makespan. Hence, the resulting transport network defines an equivalent \TDTSPTW{} instance. A forward variable fixing algorithm is defined analogously but is not used in our framework.

Like bounded labeling, the efficiency of variable fixing depends on the gap between $\ArrivalTime(\Route^*)$ and $\PricingLowerBound$. The same gap also determines how many arc-position states are removed from $\Network$. Unlike bounded labeling, however, variable fixing cannot discard a dominated label $\Mabel$ before evaluating $\CompletionBound(\Mabel)$, since $\Mabel$ may provide the tightest bound for an arc-position state. In our implementation, dominance checks are cheaper than evaluating completion bounds, making variable fixing considerably more expensive.

The sparsification method for accelerating column generation relaxes variable fixing by replacing $\RedArrivalTime(\Route^*)$ with a threshold $\SparsificationCost\in\mathbb{R}$ when computing completion values in \eqref{eq:completion:value}. When $\SparsificationCost<\RedArrivalTime(\Route^*)$, the corresponding completion bounds increase, leading to fewer arc-position combinations in the transport network. However, this reduced network may lack tours belonging to optimal solutions of $\Master$, making subsequent pricing problems heuristic rather than exact.

During the early stages of column generation, heuristic labeling is used to quickly generate columns for $\Master$ (Section~\ref{sec:pricing}). In this regime, allocating more computational effort to sparsification than to heuristic labeling is not desirable, since the generated columns may have limited impact on the final dual vector. Therefore, we further relax sparsification by ignoring the memory-inclusion condition when testing dominance, consistently with the relax-ng labeling algorithm. Additionally, relaxed sparsification is sometimes performed with $\SparsificationCost \geq \RedArrivalTime(\Route^*)$ to reduce the risk of generating a network from which no columns can be obtained in the subsequent pricing problem. Otherwise, the algorithm must revert to the original network, incurring the sparsification cost without retaining its benefits.

\section{Column generation}
\label{sec:column generation}

Algorithm~\ref{alg:column generation} presents our column-generation procedure, which implements Steps~\ref{alg:solver:CG} and~\ref{alg:solver:primal} of Algorithm~\ref{alg:solver} by improving $\Profit$ and the incumbent solution. To reduce computational effort, non-dominated labels generated during pricing are used to sparsify a working transport network $\Network_{\rm w}$, restricting subsequent pricing problems to tours deemed more promising.

\begin{algorithm}[tbh]
\caption{Column generation (high-level description)}\label{alg:column generation}
\begin{algorithmic}[1]
    \Require a transport network $\Network$, an initial set of tours $\RouteSet$, and an incumbent elementary tour $\Route^*$

    \Ensure a dual vector $\Profit$ and a set of non-dominated forward labels $\LabelSet$

    \State \Let $\Network_{\rm w} = \Network$, $\Stage = \Stage_1$, $i = 0$, and $q = -\infty$.\label{alg:column generation:init}

    \While{\textbf{true}} \label{alg:column generation:loop init}

        \State \Let $\Profit$ be an optimal dual solution of the restricted $\Master$ defined by $\RouteSet$, and $\RMPBound$ be its optimal value.

        \State \Let $\LabelSet$ be the set of labels obtained by the labeling method of $\Stage$ with input $\Profit$, and $i = i+1$.\label{alg:column generation:labeling}

        \State Add non-elementary tours in $\LabelSet$ with $\RedArrivalTime < 0$ to $\RouteSet$.

        \InlineIf{$\LabelSet$ contains elementary tours} update $\Route^*$ with an elementary tour in $\LabelSet$ minimizing $\ArrivalTime$.

        \If{$\PricingLowerBound \geq \kappa_\Stage \RMPBound$}\label{alg:column generation:do stop}

            \State Run the primal heuristic on selected tours of $\RouteSet$ to improve $\Route^*$\label{alg:column generation:primal}

            \InlineIf{$\Stage = \Stage_k$} \Output $\Profit$ and $\LabelSet$, and halt \label{alg:halt}

            \State Update $\Stage$ to the next stage if $i < \MaxPricingIterations$, and \Let $\Network_{\rm w} = \Network$, $i = 0$, and $q = -\infty$  \label{alg:column generation:reinit}


        \ElsIf{$\PricingLowerBound \geq q$}\label{alg:column generation:do sparsfication}

            \State Apply the sparsification method of $\Stage$ to $\Network_{\rm w}$ with input $\LabelSet$ and $\Profit$, and update $q$. \label{alg:column generation:sparsification}\label{alg:column generation:loop end}

        \EndIf
    \EndWhile
\end{algorithmic}
\end{algorithm}

To this end, the algorithm is organized into consecutive stages $\Stage_0,\ldots,\Stage_k$, executed throughout Loop~\ref{alg:column generation:loop init}--\ref{alg:column generation:loop end}. Each stage combines a labeling method with a sparsification strategy. Stage~$\Stage_0$ applies relax-all labeling without sparsification. Stages~$\Stage_1,\ldots,\Stage_j$ apply relax-ng labeling with relaxed sparsification, progressively increasing the threshold value $\SparsificationCost$. Stages~$\Stage_{j+1},\ldots,\Stage_{k-1}$ apply exact labeling with sparsification, again progressively increasing $\SparsificationCost$. Finally, Stage~$\Stage_k$ performs exact pricing without sparsification, computing the exact pricing lower bound $\PricingLowerBound$ and the complete set of non-dominated labels required by Algorithm~\ref{alg:solver}.

Within each stage, the algorithm proceeds through phases, each starting by resetting the working transport network $\Network_{\rm w}$ to the input network $\Network$ (Step~\ref{alg:column generation:init} or~\ref{alg:column generation:reinit}). In each phase, successive pricing problems are solved while $\Network_{\rm w}$ is progressively tightened through sparsification (Step~\ref{alg:column generation:sparsification}) until the $\Profit$-bound $\PricingLowerBound$ meets the stopping criterion controlled by $\kappa_\Stage$, which limits the tailing-off effect commonly observed in column generation.

The variable $i$, updated in Step~\ref{alg:column generation:labeling}, counts the pricing problems solved during a phase and indicates the effectiveness of the current sparsification strategy. Each phase starts by solving one pricing problem on the input transport network $\Network_{\rm w}=\Network$. Since $q$ is initialized to $-\infty$ in Steps~\ref{alg:column generation:init} and~\ref{alg:column generation:reinit}, sparsification is always applied after this first iteration. If $\Network_{\rm w}$ becomes too restrictive and only supports a small number of additional pricing iterations before the stopping criterion is met, the algorithm moves to the next stage (Step~\ref{alg:column generation:reinit}), either weakening sparsification or strengthening labeling. This mechanism avoids repeatedly restarting from the input network $\Network$ by progressively identifying configurations that maintain effective tightened networks over longer sequences of pricing iterations.

The primal heuristic tightens the completion bounds used in sparsification and variable fixing, thereby accelerating pricing through a reduced labeling search space. Since applying it to all tours added to the restricted $\Master$ may be computationally prohibitive, a limited time budget is allocated. At the end of each phase, the heuristic is applied to selected tours in $\RouteSet$ that have not been previously processed. Although high-quality incumbents are often obtained early, the heuristic remains active because further improvements are valuable when solving some of the most challenging instances.

\subsection{Additional acceleration techniques}
\label{sec:column generation:acceleration}

We implement additional acceleration techniques that provide complementary improvements with a favorable trade-off between implementation effort and performance.

First, we preserve information accumulated in the restricted $\Master$ when the state space changes. When $\Route^*$ improves, tours with $\ArrivalTime \geq \ArrivalTime(\Route^*)$ are retained in $\Master$. Similarly, after augmenting ng-memories in Step~\ref{alg:solver:DNA} of Algorithm~\ref{alg:solver}, some ng-infeasible columns are preserved by assigning them an artificially increased $\ArrivalTime$, discouraging their selection in the optimal solution. Although these columns no longer belong to $\RouteUniverse$, they provide useful dual information during the early stages of the subsequent column-generation process.

Second, we shrink time windows before starting Algorithm~\ref{alg:column generation} and whenever $\Route^*$ changes. This procedure is well established for \TSPTW{}s and time-dependent extensions \citep{AscheuerFischettiGroetschelMP2001,MonteroMendez-DiazMirandaBrontCO2017,Fontaine2024}. Its adaptation to arc-position states follows the same principles, and details are omitted. Additionally, whenever $\Route^*$ is updated, we run backward labeling on $\Network$ with $\Profit=0$ to further tighten arc-position deadlines. Although this requires exact labeling, setting $\Profit=0$ removes the profit component from Rule~\ref{rule:back-dominance}, substantially reducing the number of non-dominated labels.

Finally, we apply the ng-memory cleaner (Appendix~\ref{app:ng-memory clean}) after each sparsification or variable-fixing operation. The cleaner removes a vertex $\Zertex$ from the ng-memory associated with an arc-position state $(\Vertex\Wertex,\Position)$ whenever it proves that no cycle starting at $\Zertex$ can traverse $\Vertex\Wertex$ at position $\Position$. This reduction decreases the size of the ng-memory structure and improves the efficiency of dominance comparisons.

\section{Ng-memory augmentation (MA)}
\label{sec:memory augmentation}

Ng-memory augmentation (MA) is applied at Step~\ref{alg:solver:DNA} of Algorithm~\ref{alg:solver}, after column generation has produced a dual vector $\Profit$ yielding a strong $\Profit$-bound $\PricingLowerBound$. It relies on a \emph{cycle-forbidding} procedure applied to each tour $\Route$ belonging to a set $\RouteSet$. For each inclusion-minimal cycle $\Sequence{\Vertex_0,\ldots,\Vertex_j}$ of $\Route$ starting at position $\Position$, it inserts $\Vertex_0$ into $\Memory(\Vertex_i\Vertex_{i+1},\Position+i)$ for $i\in\Range{j-1}$, thereby removing from $\RouteUniverse$ all tours traversing the cycle.

The cycle-forbidding procedure motivates the use of arc-position states, since time windows and time-dependent travel times make the same cycle feasible at one position and infeasible at another, or lead to different completion times. This enables a more selective elimination of undesirable tours, allowing all inclusion-minimal cycles in a large set of tours to be forbidden simultaneously without excessively enlarging the ng-memory structure.

MA first applies the cycle-forbidding procedure to the tours in the optimal solution of the restricted $\Master$. Although this shrinks $\RouteUniverse$, the resulting improvement in the $\Profit$-bound $\PricingLowerBound$ after the next column-generation iteration may be limited because only a few cycles are forbidden. It therefore exploits $\Profit$ through Algorithm~\ref{alg:MA} to remove additional tours. The algorithm alternates forward and backward bounded labeling, applying the cycle-forbidding procedure to the obtained tours. It continues while the improvement in $\PricingLowerBound$ remains significant, as determined by parameter $\kappa$, which balances insufficient augmentation, causing more column-generation iterations, against excessive augmentation, causing computationally expensive pricing problems.

\begin{algorithm}[tbh]
\caption{Ng-memory augmentation (high-level description)}
\label{alg:MA}
\begin{algorithmic}[1]
    \Require a transport network $\Network$, a dual vector $\Profit$, and a set of forward labels $\LabelSet$

    \Ensure a set of forward labels $\LabelSet$

    \State \Let $d$ be the backward direction and $\kappa$ be an initial gap threshold

    \While{$d$ is backward or $\ArrivalTime(\Route^*)-\PricingLowerBound < \kappa$}\label{alg:MA:loop start}

        \State Update $\LabelSet$ to the output of the bounded labeling method in direction $d$ with input $\LabelSet$.
        \label{alg:MA:labeling}

        \State Select $\RouteSet$ as the tours with lowest $\RedArrivalTime$ among those represented by labels in $\LabelSet$.\label{alg:MA:tours}

        \State Apply the cycle-forbidding procedure to $\RouteSet$, switch direction $d$, and update $\kappa$.\label{alg:MA:forbid}

        \InlineIf{$d$ is forward} run the primal heuristic on selected tours to improve $\Route^*$.
            \label{alg:MA:primal}\label{alg:MA:loop end}

    \EndWhile

\end{algorithmic}
\end{algorithm}

In some cases, Algorithm~\ref{alg:MA} executes several iterations, substantially increasing the ng-memory structure and potentially leading to expensive pricing problems in subsequent column-generation iterations. However, this typically occurs when the gap between $\PricingLowerBound$ and $\ArrivalTime(\Route^*)$ is small and bounded labeling is computationally efficient. In such cases, Algorithm~\ref{alg:MA} further reduces the gap, allowing the exact search of Algorithm~\ref{alg:solver} (Step~\ref{alg:solver:enumerate}) to succeed and terminate the algorithm before solving any expensive pricing problem.

The primal heuristic is incorporated only after forward labeling iterations, as updating $\Route^*$ after backward labeling could invalidate completion bounds. Since backward labels are generated for the current planning horizon $\ArrivalTime(\Route^*)$, a label dominated under that horizon may become non-dominated under a tighter horizon $\Time$ due to the time-dependent nature of the problem. Consequently, these labels cannot be safely reused to compute completion bounds for horizon $\Time$. This issue does not arise for forward labeling because all forward labels start at time zero.

\section{Exact search}
\label{sec:enumeration}

Exact search (Step~\ref{alg:solver:enumerate} of Algorithm~\ref{alg:solver}) aims to obtain an optimal solution of the \TDTSPTW{} by progressively reducing the gap between the lower bound $\LowerBound$ and the incumbent solution $\Route^*$. It considers a sequence of candidate makespan thresholds $\LowerBound < \Time_0 < \ldots < \Time_k = \ArrivalTime(\Route^*)$. For $i \in \Range{k}$, a backward bounded labeling that enumerates only elementary paths is executed using the set of forward labels $\LabelSet$ obtained after Algorithm~\ref{alg:MA} to compute completion bounds.  The method uses $\Time_i$ as planning horizon and is given a limited computational budget. If this budget is exhausted, the exact search terminates without considering the remaining thresholds. Otherwise, if no tour is generated, Proposition~\ref{prop:completion bound} proves that no elementary tour in $\RouteUniverse$ satisfies $\ArrivalTime(\Route)\leq\Time_i$, and $\LowerBound$ is updated to $\Time_i$.

Conversely, if a tour is generated, then an elementary tour with makespan at most $\Time_i$ exists. The algorithm then runs a forward bounded labeling method that enumerates only elementary paths, using the backward labels $\BackLabelSet$ obtained in the previous step to compute completion bounds. The method uses the same planning horizon $\Time_i$ and runs to termination. If an elementary tour is generated, the one with minimum $\ArrivalTime$ is an optimal solution of the \TDTSPTW{}; otherwise, no elementary tour with makespan strictly smaller than $\Time_i$ exists, implying that every tour generated during the backward phase has makespan exactly $\Time_i$ and is therefore optimal.

Backward enumeration is performed before forward enumeration for the same reason discussed in Section~\ref{sec:memory augmentation}: forward labels can be used to compute completion bounds independently of the planning horizon, whereas backward labels are valid only for the horizon under which they were generated.

\section{Experimental results}
\label{sec:results}

We embedded Algorithm~\ref{alg:solver} into a branch-and-price framework \BP{} by modifying its stopping condition (Step~\ref{alg:solver:loop start}) to handle the nodes of the branch-and-bound tree. We use strong branching on the arcs of the transport network \citep{RopkePiCG2012}, following the implementation of \cite{PecinPessoaPoggiUchoaMPC2017}. The configuration of the branch-and-price method and all algorithmic components is described in Appendix~\ref{app:parametrization}.

We implement \BP{} in \CPP{} using CPLEX 12.9 as the LP solver.\footnote{Source code and supplementary results will be available upon acceptance} All experiments are performed with a one-hour time limit on a single thread of a workstation equipped with an AMD Ryzen 5 5600 CPU@$3.5$GHz and $16$GB of RAM, running Debian through WSL on Windows 11. We compare \BP{} with the methods \Lera{} \citep{Lera-RomeroMirandaBrontSoulignac2022} and \Fontaine{} \citep{FontaineDibangoyeSolnon2023}, which represent the state of the art among single-thread methods for the time-dependent TDTSPTW with loose and tight time windows, respectively. We do not compare against the time-independent solvers of \cite{TardivoMichelHoeve2026} and \cite{SoulignacCOR2026}, even on the time-independent instances, as the former is GPU-based and the latter is restricted to instances with the tightest time windows.  Extended results are reported in Appendix~\ref{app:results}.

Across our experiments, we find no systematic variation in the relative performance of \BP{}, \Lera{}, or \Fontaine{} associated with instance-generation parameters other than the number of customers and time-window width. We therefore organize the instances by these two characteristics, considering no, loose or moderate, and tight time windows in the following sections, followed by an ablation study of the sparsification mechanism.

\subsection{No time windows}
\label{sec:results:no-tw}

For the \TDTSP{} (i.e., with no time windows), we consider the following benchmark sets:
\begin{itemize}
\item \Adamo{} \citep{AdamoGhianiGuerrieroCOR2020}: 180 randomly generated \TDTSP{} instances for each $\CustCount \in {15,20,\ldots,60}$, for a total of 1800 instances.
\item \RifkiNoTW{}: 30 \TDTSP{} instances for each $\CustCount \in {10,20,\ldots,60}$ obtained from the \Rifki{} benchmark (Section~\ref{sec:results:loose}) by removing the time windows, for a total of 180 instances.
\end{itemize}

Table~\ref{tab:no-tw} reports the results of our framework for selected values of $\CustCount$. For \Adamo{}, we also report the results of \Lera{}. We do not report results for \Fontaine{} or for \Lera{} on \RifkiNoTW{}, as these instances were not considered in the respective studies.

The column $s$ gives the number of instances solved to optimality, while $t$ gives the average solution time over those instances. For our method, we also report $t_{\max}$, the maximum solution time among the instances solved to optimality, and $\times_t$ and $\times_{\max}$, which scale $t$ and $t_{\max}$, respectively, to account for the processor speed difference between our machine and that used by \Lera{} ($1.26\times$ according to \citealp{Geekbench}).

\begin{table}[htb]
    \centering\small
    \begin{tabular}{lcc ccc ccccc}
        \toprule
         &   &   & \multicolumn{2}{c}{\Lera} && \multicolumn{5}{c}{\BP} \\\cmidrule{4-5}\cmidrule{7-11}
                   &    \#   & $\CustCount$  & $s$ & $t$     && $s$ & $t$   & $t_{\max}$ & $\times_t$  & $\times_m$ \\\midrule
      \Adamo       &   180   & 20            & 178 & 220     && 180 & 1     & 2          & 1           & 2          \\
                   &   180   & 30            & 151 & 1544    && 180 & 5     & 23         & 6           & 28         \\
                   &   180   & 40            & 78  & 2891    && 180 & 26    & 52         & 33          & 65         \\
                   &   180   & 50            & --- & ---     && 180 & 187   & 2639       & ---         & ---        \\
                   &   180   & 60            & --- & ---     && 167 & 780   & 3543       & ---         & ---        \\\midrule
      \RifkiNoTW   &    30   & 30            & --- & ---     && 30  & 5     & 9          & ---         & ---        \\
                   &    30   & 40            & --- & ---     && 30  & 47    & 182        & ---         & ---        \\
                   &    30   & 50            & --- & ---     && 30  & 393   & 2003       & ---         & ---        \\
                   &    30   & 60            & --- & ---     && 21  & 1143  & 2805       & ---         & ---        \\\midrule
    \end{tabular}
    \caption{Results for instances of the \TDTSP{} (no time windows).}\label{tab:no-tw}
\end{table}

The results show a substantial performance improvement over \Lera{} on instances with no time windows. On instances solved by both methods, \BP{} is about two orders of magnitude faster on average, while also solving substantially larger instances. This behavior is consistent with the motivation for redesigning the framework: without time windows, pricing problems become harder, leading to small proportional gains after ng-memory augmentation, resulting in a less effective exact search. In contrast, jointly improving the dual vector and ng-memories, together with variable fixing, enables \BP{} to solve substantially larger instances. As we show in Section~\ref{sec:results:ablation}, sparsification helps accelerating the initial column generation phase before the first variable fixing. The results on \RifkiNoTW{} further support this behavior, with \BP{} solving most instances with up to 60 customers, including instances not considered in the original comparison.

\subsection{Loose and moderate time windows}
\label{sec:results:loose}

For the \TDTSPTW{} with loose and moderate time windows, we consider three benchmark sets designed to evaluate the effect of time-window width. Each set is derived from a common base instance using a multiplier $\TW \in {0,0.25,0.5,1}$. Given a base instance with tight time window $[\Release(\Vertex),\Deadline(\Vertex)]$ at each vertex $\Vertex \in \Range{\CustCount+1}$, the corresponding instance uses window $[\TW\Release(\Vertex),\Deadline(\Vertex)]$. Thus, $\TW=1$ preserves the original tight time windows, $\TW=0.5$ and $\TW=0.25$ yield moderate and loose time windows, respectively, whereas $\TW=0$ removes release times while retaining deadlines. The benchmark sets are:
\begin{itemize}
\item \Arigliano{} \citep{AriglianoGhianiGriecoGuerrieroPlana2019}: 300 randomly generated instances for each $\CustCount \in {15,20,30,40}$ and $\TW \in {0,0.25,0.5,1}$, for a total of 4800 instances.
\item \Rifki{} \citep{Fontaine2024}: 150 instances for each $\CustCount \in {10,20,\ldots,60}$ and $\TW \in {0,0.25,0.5,1}$, for a total of 3600 instances. The benchmark is based on that of \cite{RifkiChiabautSolnon2020}, which was generated using realistic traffic simulation with real data from the city of Lyon. Travel times are taken from shortest paths for departure times separated by six minutes.

\item \RifkiConst \citep{Fontaine2024}: 60 \TSPTW{} instances for each $\CustCount \in {20,30,40}$ and $\TW \in {0,0.25,0.5,1}$, obtained from \Rifki{} by removing time dependency, for a total of 720 instances.
\end{itemize}

We also consider the \Pesant{} \citep{PesantGendreauPotvinRousseau1998} and \PotvinBengio{} \citep{PotvinBengio1996} \TSPTW{} benchmarks, obtained by extracting vehicle routes from solutions to Solomon's RC2 vehicle routing instances with time windows. Although these benchmarks contain at most 45 customers, they are considered difficult, with one instance still open according to \cite{Lopez-IbanezBlum2023}. We exclude the other benchmark sets collected by \cite{Lopez-IbanezBlum2023} because they are all easily solved by \cite{SoulignacCOR2026}.

\begin{table}[htb]
    \centering\small
    \begin{tabular}{lccc ccc ccc ccccc}
        \toprule
         &   &   &    & \multicolumn{2}{c}{\Lera} && \multicolumn{2}{c}{\Fontaine} && \multicolumn{5}{c}{\BP} \\\cmidrule{5-6}\cmidrule{8-9}\cmidrule{11-15}
                   &    \#   & $\CustCount$  & $\TW$ & $s$ & $t$     && $s$      & $t$     && $s$ & $t$   & $t_{\max}$ & $\times_t$  & $\times_m$ \\\midrule
      \Arigliano   &    300  & 30            & 0     & 300 & 1788    && 300      & 496     && 300 & 5     & 34         & 13          & 84         \\
                   &    300  & 40            & 0     & 126 & 2778    && 0        & ---     && 300 & 58    & 867        & 146         & 2177       \\\midrule
      \Rifki       &    150  & 30            & 0     & 149 & 2176    && 149      & 397     && 150 & 9     & 101        & 22          & 252        \\
                   &    150  & 40            & 0     & 11  & 2902    && 0        & ---     && 150 & 131   & 2010       & 329         & 5044       \\
                   &    150  & 50            & 0     & --- & ---     && ---      & ---     && 137 & 817   & 3582       & ---         & ---        \\
                   &    150  & 60            & 0     & --- & ---     && ---      & ---     && 64  & 1669  & 3542       & ---         & ---        \\\midrule
      \RifkiConst  &    60   & 30            & 0     & 60  & 1376    && 60       & 129     && 60  & 4     & 20         & 11          & 49         \\
                   &    60   & 40            & 0     & 34  & 2837    && 5        & 2011    && 60  & 69    & 653        & 174         & 1639       \\\midrule
      \Arigliano   &    300  & 30            & 0.25  & 300 & 1084    && 300      & 145     && 300 & 4     & 27         & 11          & 68         \\
                   &    300  & 40            & 0.25  & 244 & 2593    && 35       & 2444    && 300 & 40    & 425        & 100         & 1068       \\\midrule
      \Rifki       &    150  & 30            & 0.25  & 150 & 1503    && 150      & 68      && 150 & 4     & 15         & 11          & 37         \\
                   &    150  & 40            & 0.25  & 132 & 2744    && 27       & 1950    && 150 & 33    & 381        & 83          & 955        \\
                   &    150  & 50            & 0.25  & --- & ---     && ---      & ---     && 149 & 202   & 1025       & ---         & ---        \\
                   &    150  & 60            & 0.25  & --- & ---     && ---      & ---     && 142 & 989   & 3479       & ---         & ---        \\\midrule
      \RifkiConst  &    60   & 30            & 0.25  & 60  & 643     && 60       & 29      && 60  & 3     & 31         & 8           & 77         \\
                   &    60   & 40            & 0.25  & 58  & 2318    && 47       & 1221    && 60  & 34    & 188        & 86          & 473        \\\midrule
      \Arigliano   &    300  & 30            & 0.5   & 300 & 389     && 300      & 5       && 300 & 2     & 6          & 5           & 14         \\
                   &    300  & 40            & 0.5   & 300 & 1837    && 280      & 645     && 300 & 16    & 133        & 40          & 334        \\\midrule
      \Rifki       &    150  & 30            & 0.5   & 150 & 431     && 150      & 1       && 150 & 2     & 6          & 6           & 14         \\
                   &    150  & 40            & 0.5   & 149 & 1450    && 150      & 35      && 150 & 10    & 35         & 25          & 88         \\
                   &    150  & 50            & 0.5   & --- & ---     && ---      & ---     && 150 & 49    & 204        & ---         & ---        \\
                   &    150  & 60            & 0.5   & --- & ---     && ---      & ---     && 150 & 258   & 2576       & ---         & ---        \\\midrule
      \RifkiConst  &    60   & 30            & 0.5   & 60  & 129     && 60       & 0       && 60  & 1     & 3          & 3           & 7          \\
                   &    60   & 40            & 0.5   & 60  & 888     && 60       & 26      && 60  & 8     & 21         & 19          & 53         \\\midrule
      \Pesant      &    27   & 19--44        & ---   & 22  & 203     && 27       & 66      && 27  & 2     & 18         & 4           & 46         \\\midrule
      \PotvinBengio&    30   & 3--45         & ---   & 25  & 160     && 29       & 83      && 30  & 3     & 53         & 9           & 132        \\\bottomrule
  \end{tabular}
  \caption{Results for instances of the \TDTSPTW{} with loose and moderate time windows.}\label{tab:loose}
\end{table}

Table~\ref{tab:loose} reports the results of \BP{}, \Lera{}, and \Fontaine{} for selected values of $\CustCount$; the results for \Lera{} and \Fontaine{} are taken from \cite{Fontaine2024}, and our processor is approximately $2.51$ times faster according to \cite{Geekbench}. Missing entries correspond to unreported results. The results show that \BP{} solves all instances with $\CustCount\leq40$ and most instances with $\CustCount\in\{50,60\}$. As observed for \Lera{} and \Fontaine{}, its performance generally improves as time windows become tighter. Nevertheless, \BP{} remains competitive with, and overall slightly faster than, \Fontaine{} for $\TW=0.5$, while its advantage is much larger for $\TW\in\{0,0.25\}$.

Interestingly, \BP{} performs slightly worse for $\TW=0$ than on instances with no time windows, showing that adding deadlines alone can make the problem harder for our framework. This does not contradict Remark~\ref{rem:time-windows}, since the generalization there relies on potentially discontinuous travel-time functions, whereas those in our benchmarks are continuous.  Table~\ref{tab:ablation:first} in Section~\ref{sec:results:ablation} indicates that, for \BP{}, the gap between $\ArrivalTime(\Route^*)$ and $\LowerBound$ after the initial column generation is larger for $\TW=0$. Consequently, although pricing is harder without time windows during the initial column generation, the first variable-fixing step removes more arc-position states, and fewer solver iterations are required to close the gap. This difference is not observed for \Lera{}, which does not depend on variable fixing.

For $\TW=0.5$, the performance gap with \Fontaine{} is substantially smaller. This difference reflects the solving strategies of the two frameworks. \Fontaine{} relies on a single exact best-first search, guided by alternative completion bounds, that becomes more effective as the time windows tighten. In contrast, \BP{} invests computational effort in obtaining strong dual information, which it then uses to progressively remove arc-position states and reduce the optimality gap. This strategy is particularly beneficial for loose time windows, but becomes less attractive as the windows tighten, when pricing over the larger set of ng-feasible tours can be more expensive than directly solving the increasingly well-structured exact problem. A similar trade-off explains \Lera{}'s behavior, whose first two phases obtain backward labels that provide completion bounds for the exact search in the final phase.

Finally, \BP{} solves all instances in the \Pesant{} and \PotvinBengio{} benchmarks in few seconds, including the one reported open by \cite{Lopez-IbanezBlum2023}.

\subsection{Tight time windows}
\label{sec:results:tight}

To study tight time windows, we consider the \Vu{} benchmark \citep{VuHewittBolandSavelsberghTS2020}. These instances extend the generation procedure of \Arigliano{} to larger numbers of customers, but restrict the time-window widths to $w \in \{40,60,80,100,120,150,180\}$, which are tighter than those used in \Arigliano{}.

\begin{table}[htb]
    \centering\small
    \begin{tabular}{lccc ccc ccc ccccc}
        \toprule
         &   &   &    & \multicolumn{2}{c}{\Lera} && \multicolumn{2}{c}{\Fontaine} && \multicolumn{5}{c}{\BP} \\\cmidrule{5-6}\cmidrule{8-9}\cmidrule{11-15}
                   &    \#   & $\CustCount$  & $w$       & $s$ & $t$     && $s$      & $t$     && $s$ & $t$   & $t_{\max}$ & $\times_t$  & $\times_m$ \\\midrule
      \Vu          
                   &    120  & 100           & $\leq 80$ & 120 & 58      && 120      & 0       && 120 & 31    & 74         & 79          & 186        \\
                   &    40   & 100           & 100       & 40  & 219     && 40       & 1       && 40  & 47    & 113        & 117         & 284        \\
                   &    40   & 100           & 120       & 40  & 366     && 40       & 5       && 40  & 53    & 327        & 134         & 821        \\
                   &    40   & 100           & 150       & 38  & 723     && 40       & 79      && 40  & 54    & 169        & 135         & 425        \\\bottomrule
  \end{tabular}
  \caption{Results for instances of the \TDTSPTW{} with tight time windows for $\CustCount = 100$.}\label{tab:tight}
\end{table}

Table~\ref{tab:tight} shows that \BP{} substantially improves upon \Lera{}, but \Fontaine{} is considerably faster, especially in the most restricted regimes. Results reported in the supplementary material explain this difference: the arc-position representation makes \BP{} spend non-negligible time shrinking time windows, whereas \Fontaine{} shrinks the time windows associated with the vertices almost instantly. \BP{} then incurs additional effort to obtain a strong dual vector over an already tightly constrained instance. This overhead becomes increasingly relevant as $\CustCount$ grows, as observed in unreported experiments on the instances collected by \cite{Lopez-IbanezBlum2023} with $\CustCount>50$: in some cases, the initial column generation of \BP{} does not converge within the one-hour limit, whereas \Fontaine{} and \cite{SoulignacCOR2026} solve the same instances in a few seconds. These results indicate that computing a strong $\Profit$-bound over the relaxed space of ng-feasible tours is not attractive for tight time windows, where A$^*$-like exact search can exploit the temporal restrictions directly. This suggests combining A$^*$-like exact search and column-generation based methods according to the time-window regime, as proposed in our previous work \citep{SoulignacCOR2026}.

\subsection{Ablation studies}
\label{sec:results:ablation}

In a first ablation study, we compare three configurations of the framework during the first iteration of Algorithm~\ref{alg:solver} (Solver). The full configuration \BP{} is the one used in all previous experiments. In the intermediate configuration \Sparse{}, we disable the primal heuristic, so that the upper bound is obtained solely from tours generated during pricing. Finally, in the base configuration \Base{}, we further disable sparsification by restricting column generation to three stages, corresponding to the relax-all, relax-ng, and exact configurations. We consider $\CustCount=50$ customers on the \RifkiNoTW{} and \Rifki{} ($\TW=0$) benchmarks, focusing on loose-time-window regimes. To ensure a fair comparison, we perform exactly four iterations in each direction of Loop~\ref{alg:MA:loop start}--\ref{alg:MA:loop end} of Algorithm~\ref{alg:MA} (Ng-memory augmentation) in all configurations. We also exclude two instances that required less than 5 seconds for Ng-memory augmentation in at least one configuration.

\begin{figure}
 \centering
 \begin{tabular}{c@{ }c@{ }c@{ }c@{ }c}
   \includegraphics[scale=.35]{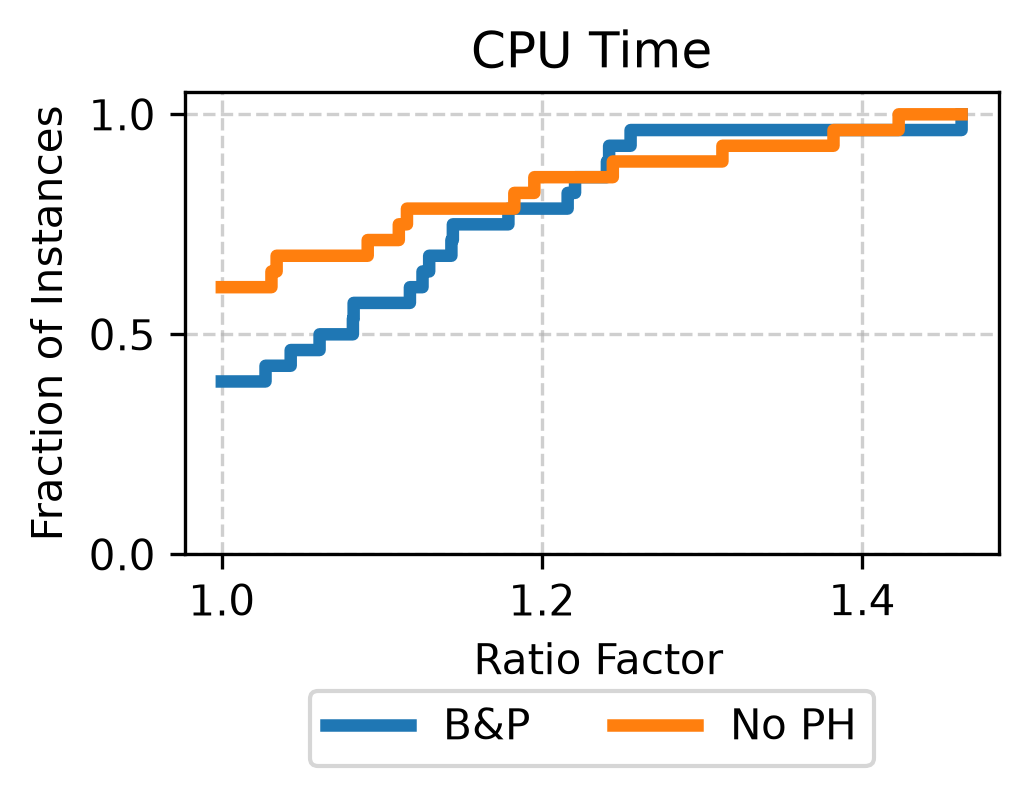} & \includegraphics[scale=.35]{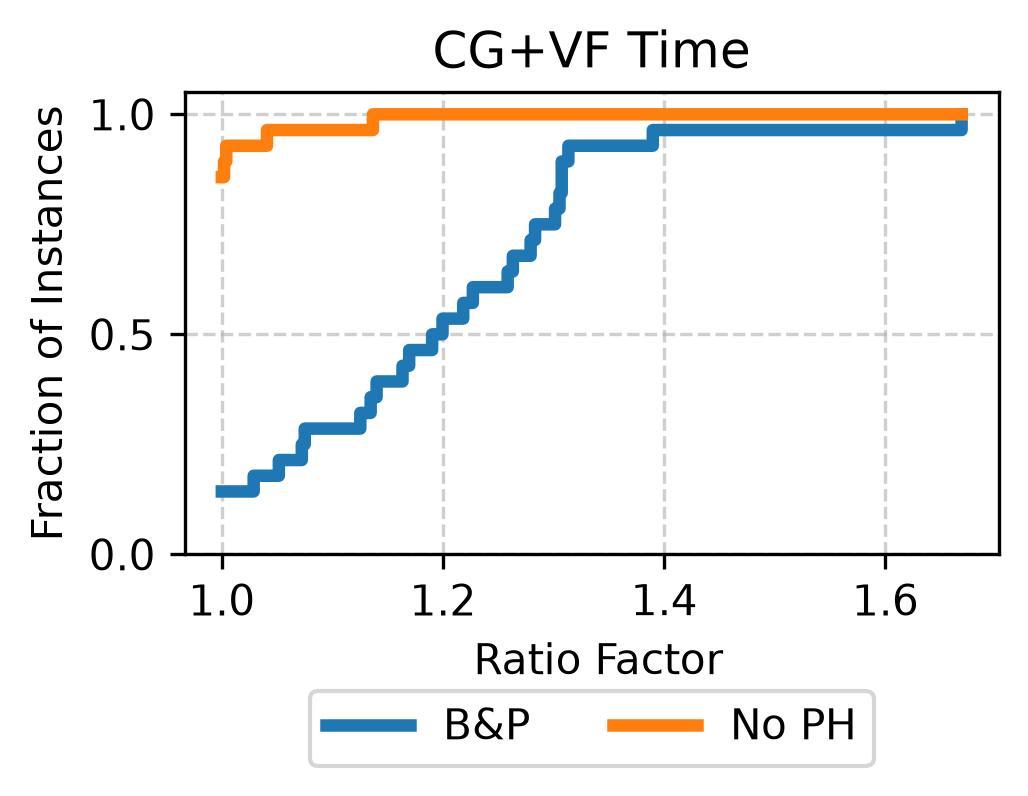} & \includegraphics[scale=.35]{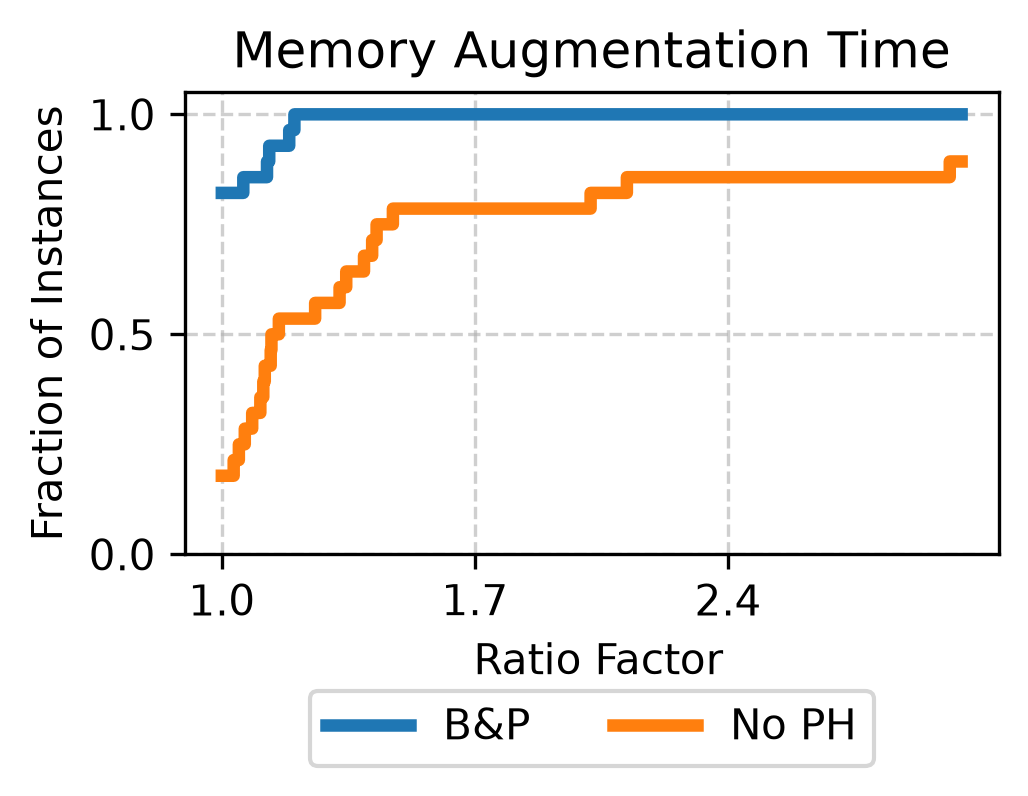} & \includegraphics[scale=.35]{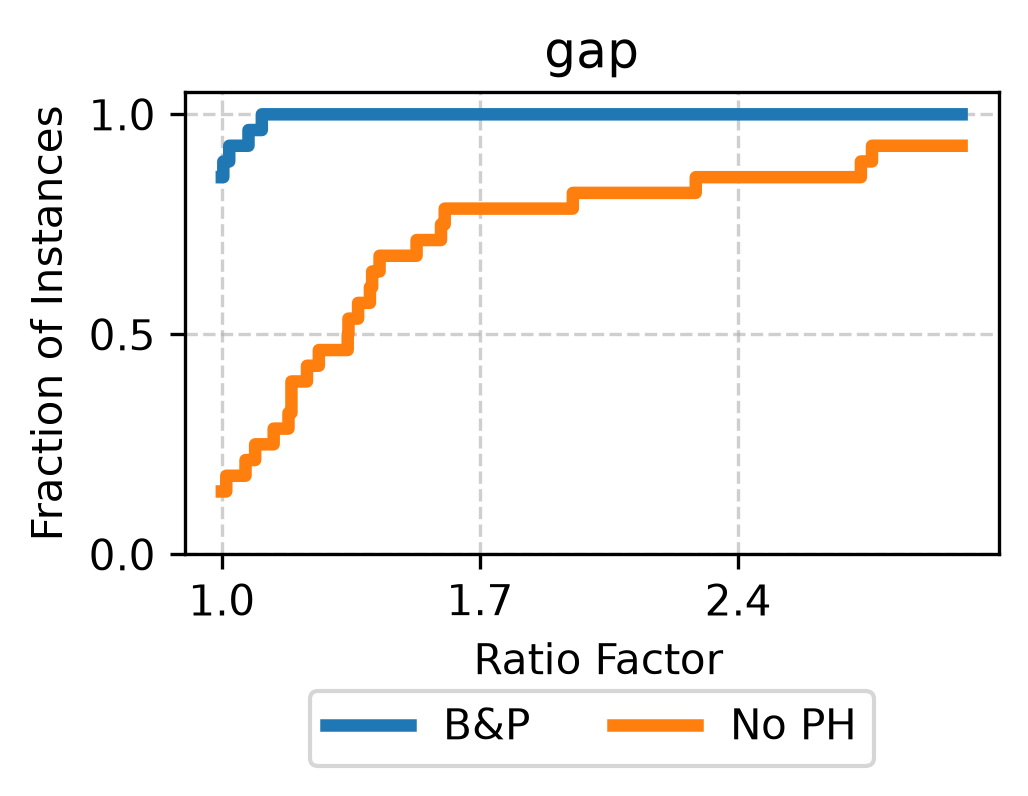} & \includegraphics[scale=.35]{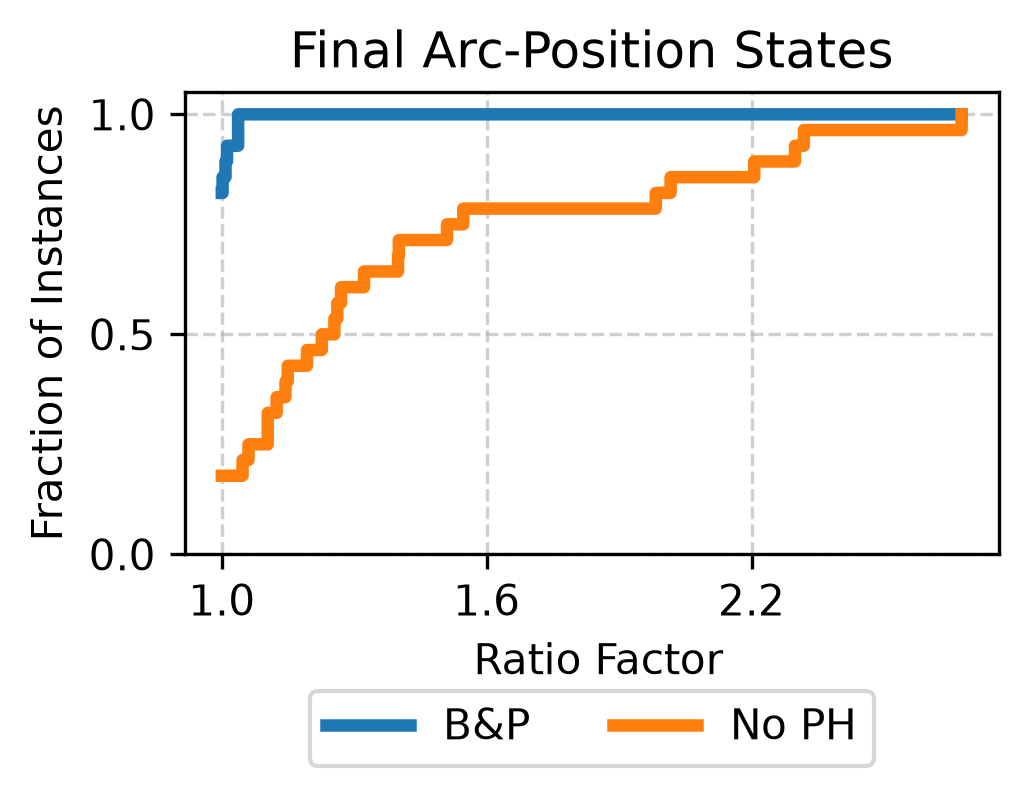} \\
   \includegraphics[scale=.35]{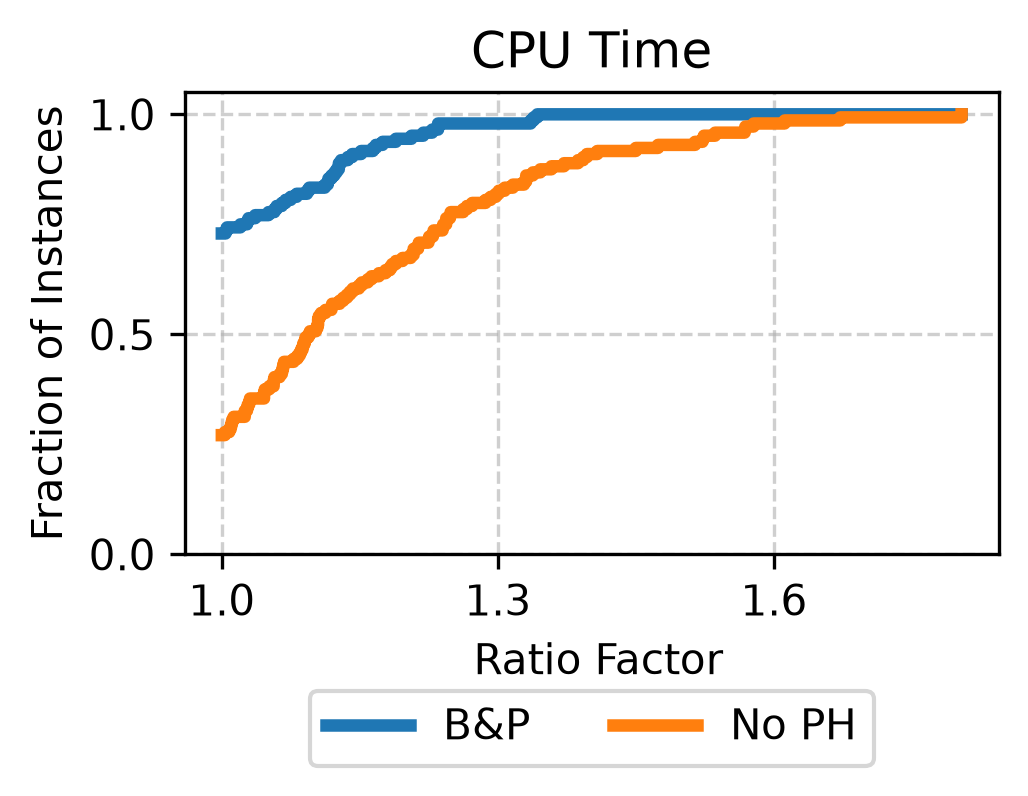} & \includegraphics[scale=.35]{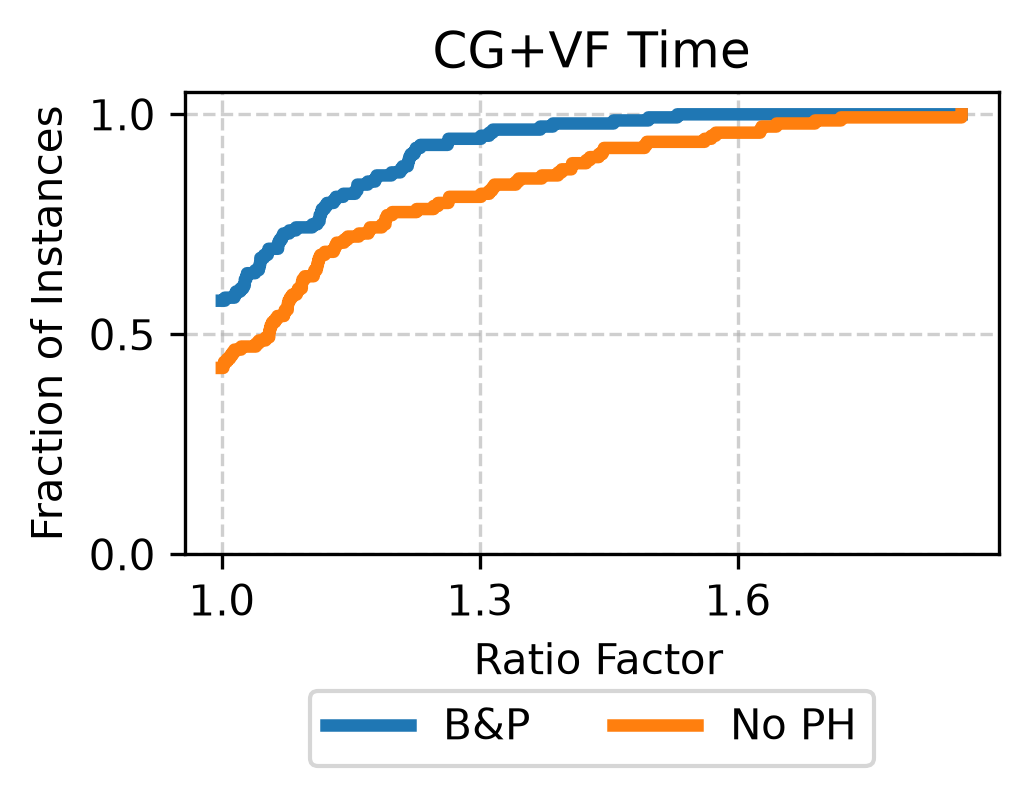} & \includegraphics[scale=.35]{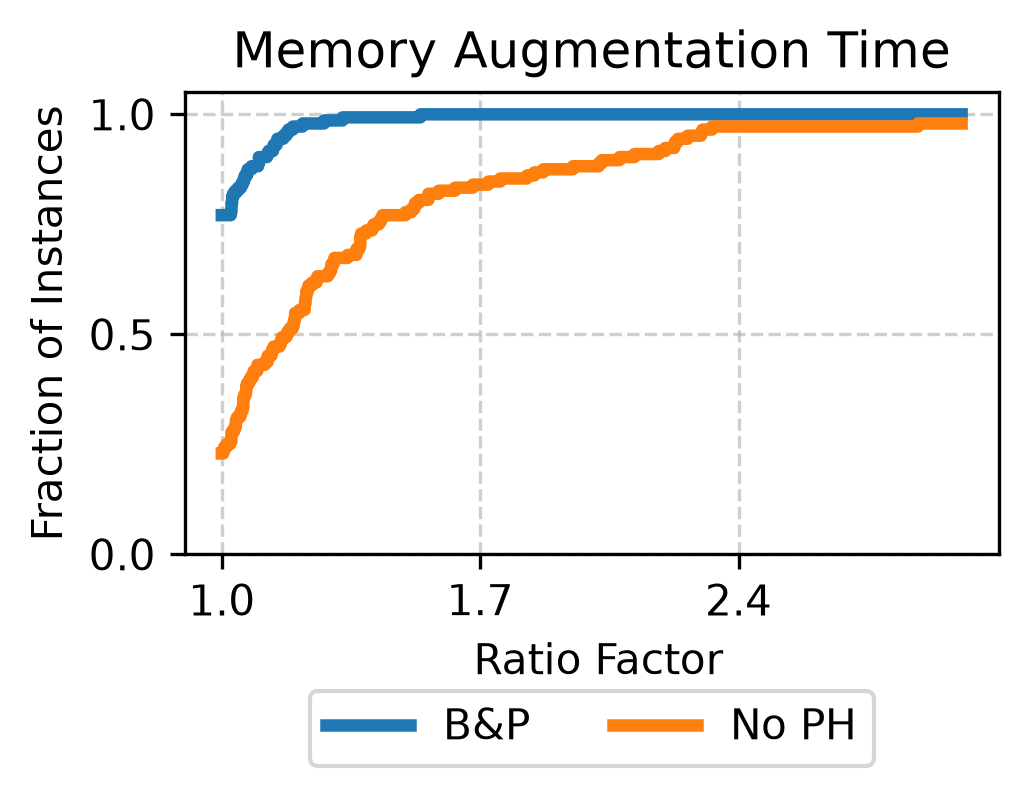} & \includegraphics[scale=.35]{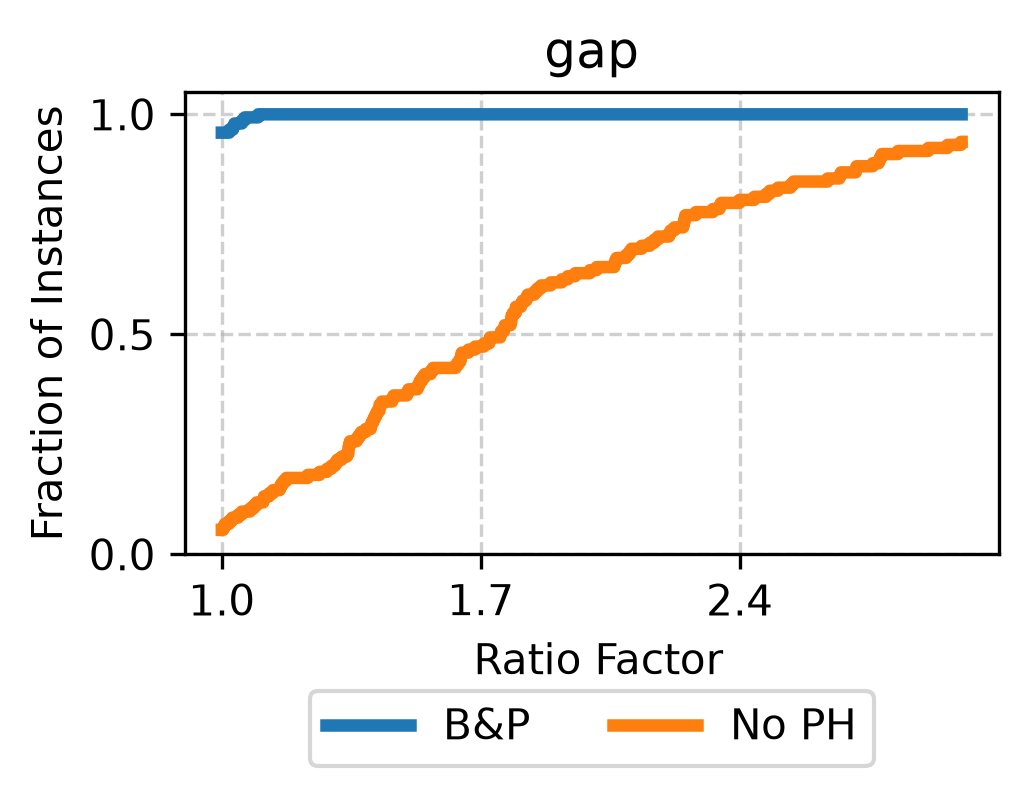} & \includegraphics[scale=.35]{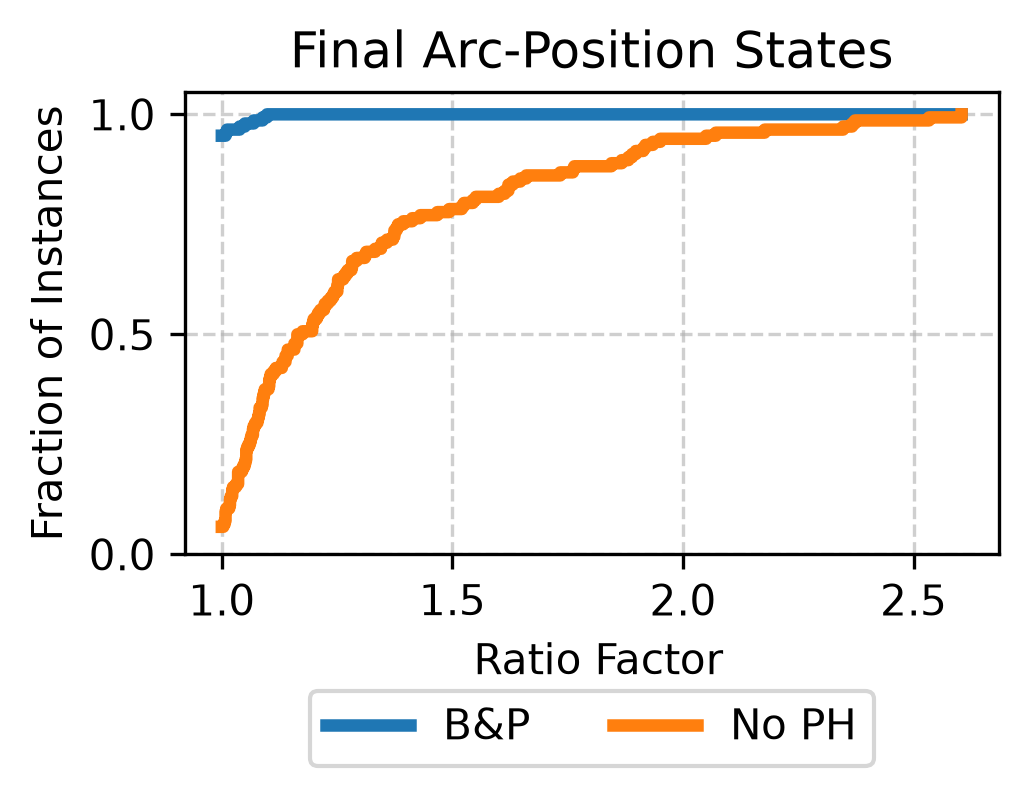}
 \end{tabular}
 \caption{Dolan-Moré performance profiles for \RifkiNoTW{} (above) and \Rifki{} (below) on the first iteration of Solver~\ref{alg:solver}.}
 \label{fig:Dolan-More}
\end{figure}

Figure~\ref{fig:Dolan-More} depicts Dolan--Moré performance profiles comparing \Sparse{} and \BP{}, while Table~\ref{tab:ablation:first} reports aggregated results. The columns labeled avg.\ report the average CPU time (CPU), column generation plus variable fixing time (CG+VF, Steps~\ref{alg:solver:CG}--\ref{alg:solver:VF} of Algorithm~\ref{alg:solver}), Ng-memory augmentation time (MA, Step~\ref{alg:solver:DNA} of Algorithm~\ref{alg:solver}), gap percentage (Gap\%), and number of feasible arc-position states (States) after the variable fixing step. The Gap\% is computed as $100(\ArrivalTime(\Route^*)-\LowerBound)/\ArrivalTime(\Route^*)$, where $\Route^*$ and $\LowerBound$ are the incumbent and lower bound obtained by each configuration. The rel.\ columns report the percentage of the \Base{} value saved by each configuration, computed as $100(\text{value}_{\Base}-\text{value}_{X})/\text{value}_{\Base}$. Larger rel.\ values indicate greater improvements over \Base{}.

\begin{table}[htb]
    \centering\small
    \begin{tabular}{ll rrr rrr rrr rrr rr}
        \toprule
                     &       & \multicolumn{2}{c}{CPU} && \multicolumn{2}{c}{CG+VF} && \multicolumn{2}{c}{MA} && \multicolumn{2}{c}{Gap\%} && \multicolumn{2}{c}{States} \\\cmidrule{3-4}\cmidrule{6-7}\cmidrule{9-10}\cmidrule{12-13}\cmidrule{15-16}
                     &                & avg.   & rel. && avg.   & rel. && avg.  & rel. && avg.  & rel. && avg.  & rel. \\\midrule
          \RifkiNoTW & \Base{}        & 239    & 0    && 179    & 0    && 60    & 0    && 6.58  & 0    && 81130 & 0    \\
                     & \Sparse{}      & 129    & 46\% && 80     & 55\% && 49    & 18\% && 5.17  & 21\% && 68187 & 16\% \\
                     & \BP            & 133    & 44\% && 95     & 47\% && 38    & 36\% && 3.64  & 45\% && 51075 & 38\% \\\midrule
  \Rifki{} ($\TW=0$) & \Base{}        & 215    & 0    && 160    & 0    && 55    & 0    && 16.62 & 0    && 95640 & 0    \\
                     & \Sparse{}      & 167    & 22\% && 115    & 28\% && 52    & 5\%  && 15.25 & 8\%  && 90610 & 5\%  \\
                     & \BP            & 153    & 29\% && 108    & 32\% && 45    & 19\% &&  7.67 & 54\% && 74021 & 23\% \\\midrule
    \end{tabular}
    \caption{Effects of sparsification and primal heuristic on the first iteration of Solver~\ref{alg:solver}.}\label{tab:ablation:first}
\end{table}

The aggregated results in Table~\ref{tab:ablation:first} show that sparsification reduces the average time required by the first iteration by roughly one-fourth to one-half. Most of this reduction comes from column generation. Interestingly, sparsification also substantially reduces the average final gap when the primal heuristic is disabled. This improvement could be related to the smaller search space induced by sparsification, which may increase the likelihood that an elementary tour is among the limited number of tours obtained in each pricing iteration. However, sparsification may also eliminate states leading to elementary tours, so this interpretation is only a speculation.

When comparing \Sparse{} and \BP{}, the former is slightly faster on average on the instances without time windows but somewhat slower on those with loose time windows. Overall, the additional cost of the primal heuristic is justified by the substantially smaller gaps and number of arc-position states obtained by \BP{} in nearly all instances, providing a stronger starting point for subsequent iterations.

In our second ablation study, we assess the overall impact of sparsification while retaining the primal heuristic. We compare \BP{} with \PHOnly{}, which extends \Base{} by allocating the primal heuristic the same budget relative to pricing as in \BP{}. Again, we consider $\CustCount=50$ on the \RifkiNoTW{} and \Rifki{} ($\TW=0$) benchmarks. Table~\ref{tab:ablation:total} reports the number of solved instances ($s$), and the average CPU time ($t$), and pricing, sparsification, and variable fixing time ($p$) over the instances solved by \PHOnly{}, and the gap percentage over the instances not solved by \PHOnly{} ($g$).

\begin{table}[htb]
    \centering\small
    \begin{tabular}{l ccccc cccc}
        \toprule
         & \multicolumn{4}{c}{\PHOnly} && \multicolumn{4}{c}{\BP} \\\cmidrule{2-5}\cmidrule{7-10}
                              & $s$ & $t$  & $p$  & $g$   && $s$ & $t$  & $p$   & $g$  \\\midrule
        \RifkiNoTW            & 29  & 544  & 302  & 3.28  && 30  & 350  & 144   & 0    \\
        \Rifki{} ($\TW = 0$)  & 131 & 1022 & 571  & 4.38  && 137 & 817  & 343   & 2.57 \\\bottomrule
    \end{tabular}
    \caption{End-to-end effects of sparsification.}\label{tab:ablation:total}
\end{table}

\begin{figure}
 \centering
 \begin{tabular}{c@{ }c@{ }c@{ }c}
   \includegraphics[scale=.35]{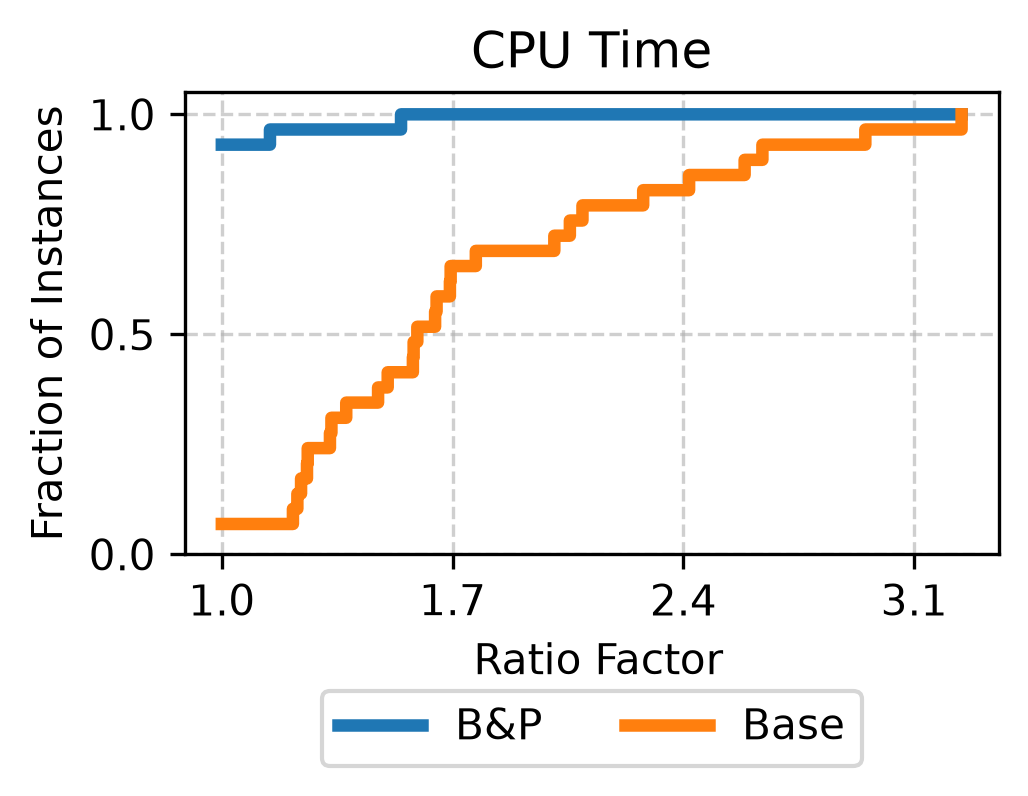} & \includegraphics[scale=.35]{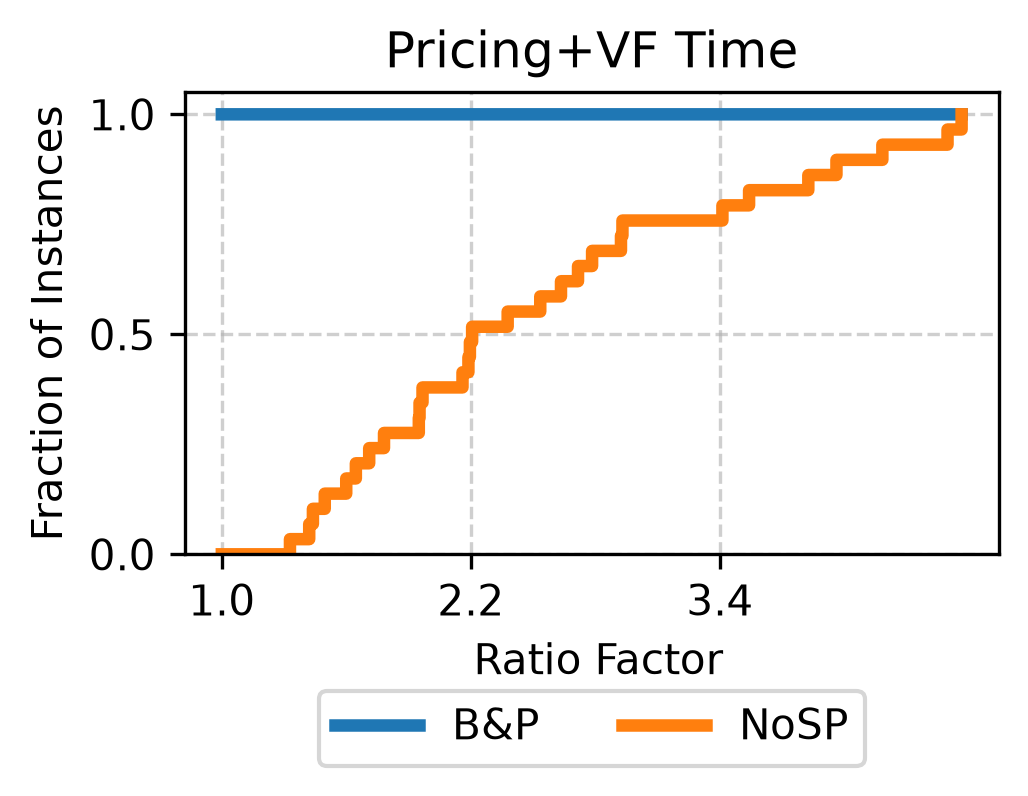} & \includegraphics[scale=.35]{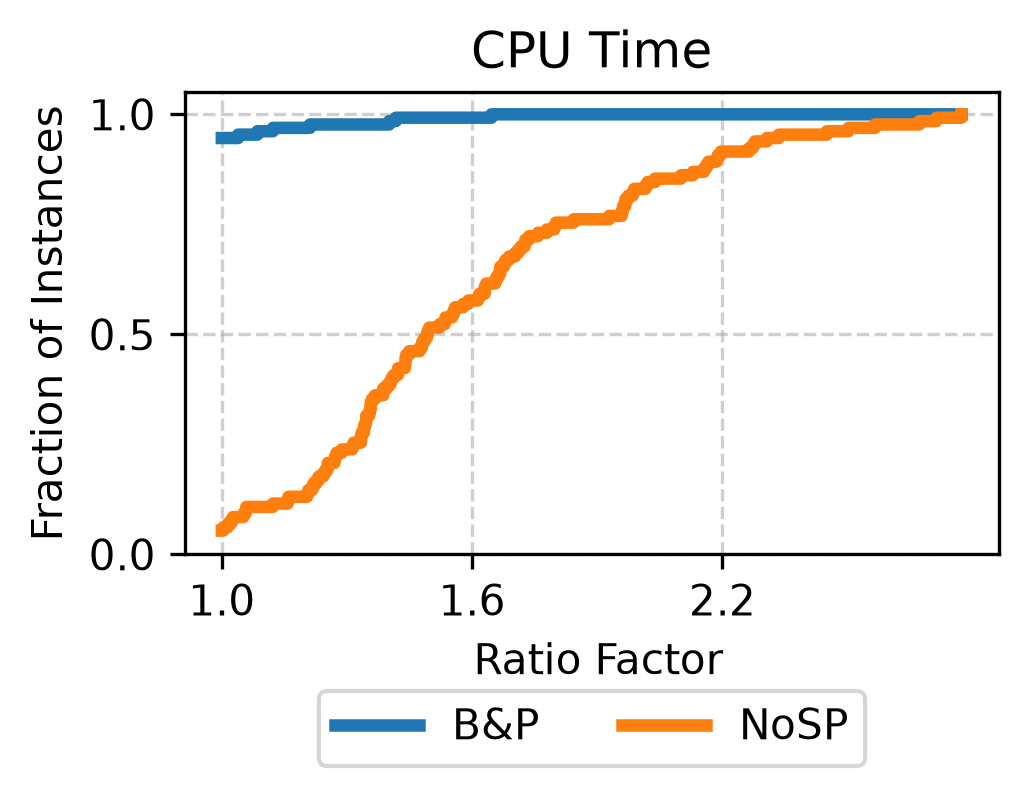} & \includegraphics[scale=.35]{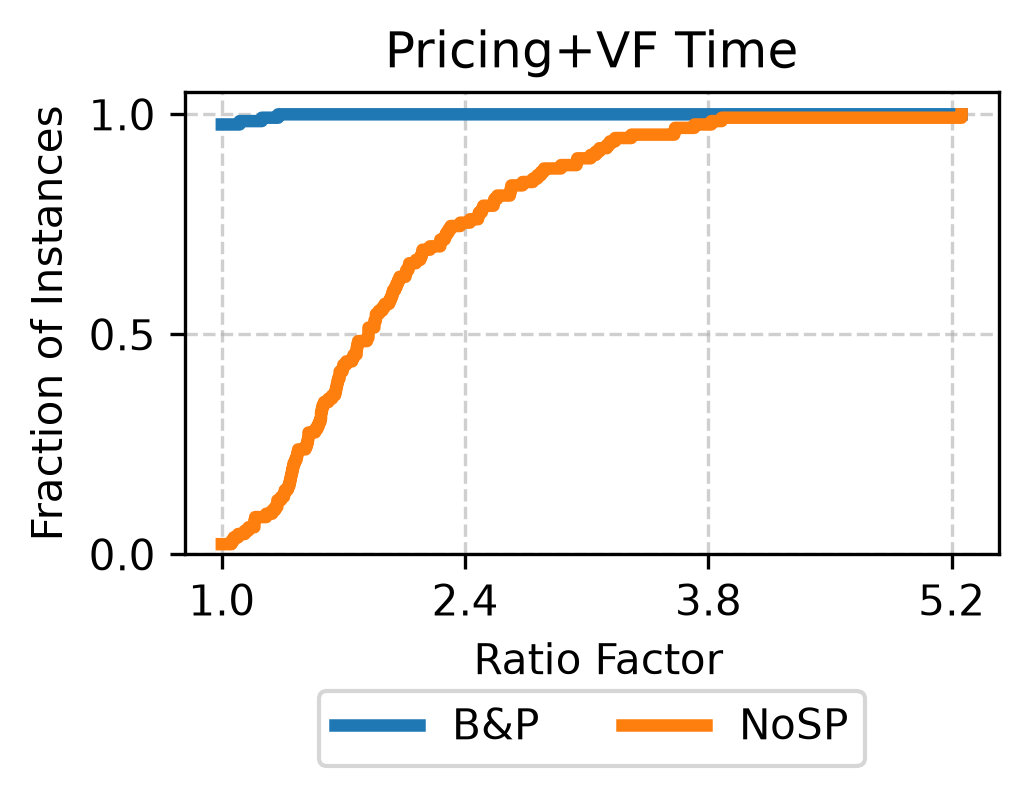}
 \end{tabular}
 \caption{Dolan-Moré performance profiles for \RifkiNoTW{} (left) and \Rifki{} (right) on end-to-end experiments.}
 \label{fig:Dolan-More:end2end}
\end{figure}

The aggregated results show that \BP{} solves more instances and requires less time on all instances solved by \PHOnly{}. The savings come almost exclusively from pricing, sparsification, and variable fixing, while the remaining computational effort is similar for both configurations. Figure~\ref{fig:Dolan-More:end2end} complements Table~\ref{tab:ablation:total}, showing that \PHOnly{} requires about 1.8 times (loose time windows) and 2.2 times (no time windows) the time spent by \BP{} on pricing, sparsification, and variable fixing on half of the instances, with ratios reaching up to 4.5 and 5.2, respectively. These ratios are computed over the instances solved by \PHOnly{}, which are all solved by \BP{} as well, and thus do not reflect the additional instances solved only by \BP{}.

\section{Conclusions}
\label{sec:conclusions}

We developed a new exact framework for the \TDTSPTW{} that extends the solving capabilities of dynamic-programming-based methods to instances with loose or absent time windows. The framework substantially improves upon the method of \cite{Lera-RomeroMirandaBrontSoulignac2022}, solving all instances with up to 40 customers and a large fraction of those with 60 customers within the considered time limit, including \TDTSP{} and \TSPTW{} instances that could not be solved by previous methods. These results raise a broader question about the most appropriate algorithmic paradigm as temporal restrictions and time dependence are progressively removed. In both time-dependent routing without time windows and time-independent routing with loose time windows, dynamic-programming-based methods currently appear to have a substantial advantage over other exact approaches. In contrast, for the classical \TSP{}, where both time dependence and time-window restrictions are absent, mathematical-programming-based methods have long been highly effective. Whether this change in the relative performance of algorithmic paradigms reflects an intrinsic effect of time dependence and temporal restrictions or simply the current state of the competing methods remains an open question. Addressing the time-dependent case requires considering \TDTSP{} instances whose travel-time functions do not implicitly impose restrictions on the feasible arrival times at customers, since, as discussed in Remark~\ref{rem:time-windows}, temporal restrictions can otherwise be encoded directly into the travel-time functions.

The computational results also suggest that no single dynamic-programming-based strategy is likely to be uniformly effective across different regimes of time-window tightness. Decision diagrams and informed search methods that solve the exact problem directly, such as \cite{FontaineDibangoyeSolnon2023,SoulignacCOR2026,TardivoMichelHoeve2026}, benefit from tight time windows because they substantially restrict the feasible state space, whereas the additional effort required by column generation and ng-relaxations becomes increasingly worthwhile as time windows become looser. Identifying the transition between these regimes, and developing an effective instance-dependent algorithm selection strategy, remains an open question.

The effectiveness of the proposed framework ultimately relies on balancing the bounding strength of $\Profit$ against the computational effort required to obtain it. Column generation and ng-memory augmentation progressively improve the $\Profit$-bound, while the primal heuristic improves the incumbent and, consequently, tightens the completion bounds used to restrict the network. In particular, completion-bound-based sparsification substantially reduces the cost of the initial pricing problems, before variable fixing can effectively restrict the network, allowing the subsequent bound-strengthening mechanisms to become effective. The resulting interaction between primal improvements, $\Profit$-bounds, and selective network reduction at the arc-position level is key to preventing stronger ng-relaxations from causing a prohibitive growth in the pricing state space. This balance appears to be essential for extending exact solution methods to the weakly constrained regime of the \TDTSPTW{}.

\section*{Declaration of generative AI and AI-assisted technologies in the manuscript preparation process}

During the preparation of this work, the author used OpenAI's ChatGPT (Luna 5.6) to assist with the editing and refinement of the manuscript, including improvements to sentence structure, organization, clarity, and scientific writing in English. The tool was also used to assist with the presentation of methodological details and experimental results. The author provided the underlying scientific content, methodological decisions, experimental results, and interpretations. After using the tool, the author reviewed and edited the content as needed and takes full responsibility for the content of the article.


\appendix

\section{Primal heuristic}
\label{app:primal}

The primal heuristic is used to generate improved elementary tours from potentially non-elementary tours obtained during the solution process. It combines an elementary conversion procedure, a destroy-and-repair heuristic, and a local search phase based on classical TSP neighborhood operators, namely swap, 2-opt, and shift moves. These components are complementary: destroy-and-repair is particularly useful when time windows strongly restrict set of feasible tours, whereas local search becomes increasingly relevant as these restrictions are relaxed, especially in the absence of time windows.

Given a tour $\Route \in \RouteUniverse$, the heuristic first removes repeated visits, keeping only the first occurrence of each customer, and reinserts the missing customers using the greedy repair procedure. It then applies destroy-and-repair by iteratively removing a small number of consecutive customers and reinserting them in a random order. Insertions are evaluated lexicographically, first minimizing the total time-window violation and then the completion time. The heuristic is repeated until a fixed number of iterations without improvement is reached, after which the local search phase is applied.

\section{Ng-memory cleaning}
\label{app:ng-memory clean}

The \emph{clean-ng method} reconstructs the ng-memory structure $\Memory$ of a transport network $\Network$ while preserving the ng-feasibility status of all tours. For each vertex $\Zertex$, it removes $\Zertex$ from $\Memory(\Vertex\Wertex,\Position)$ whenever, after the removal, no $\Memory$-infeasible tour contains a cycle starting and ending at $\Zertex$ that traverses arc $\Vertex\Wertex$ at position $\Position$.

Starting from an empty memory structure $\Memory'$, the method processes one customer at a time. For each $\Zertex \in \Range{1,\CustCount}$, it builds a position-expanded digraph $\Network_\Zertex$ containing a vertex $\Vertex_\Position$ for every vertex $\Vertex \in \Range{\CustCount+1}$ and position $\Position \in \Range{\CustCount+1}$. For $\Wertex \in \Range{\CustCount+1}$ and $\Position \leq \CustCount$, $\Network_\Zertex$ contains an arc from $\Vertex_\Position$ to $\Wertex_{\Position+1}$ if and only if $\Zertex \in \Memory(\Vertex\Wertex,\Position)$. Consequently, every $\Memory$-infeasible tour in $\Network$ visiting $\Zertex$ at positions $i$ and $j$ is represented by a path from $\Zertex_i$ to $\Zertex_j$ in $\Network_\Zertex$. Therefore, $\Zertex$ can be safely omitted from $\Memory'(\Vertex\Wertex,\Position)$ if $\Network_\Zertex$ contains no feasible paths traversing arc $\Vertex_\Position\Wertex_{\Position+1}$.

Each position-expanded vertex $\Vertex_\Position$ is associated with the time window
$[\Release(\Vertex_\Position),\Deadline(\Vertex,\Position)]$, where
\begin{align*}
 \Release(\Vertex_\Position) &= \min\{\Release(\Vertex\Wertex,\Position) \mid \Wertex \in \Range{\CustCount+1}\},\\
 \Deadline(\Vertex_\Position) &= \max\{\Deadline(\Wertex\Vertex,\Position-1) \mid \Wertex \in \Range{\CustCount+1}\}.
\end{align*}
For every path in $\Network_\Zertex$ from $\Zertex_\Position$ to $\Zertex_{\Position+k}$, earliest arrival times and latest feasible departure times are computed according to~\eqref{eq:earliest-arrival-time} and~\eqref{eq:latest-departure-time}, using the arc-position time windows of $\Network$. The only difference is that the base case of~\eqref{eq:earliest-arrival-time} initializes the recursion with $\Release(\Zertex_\Position)$ instead of the departure time $0$ at the start depot, whereas the base case of~\eqref{eq:latest-departure-time} initializes the recursion with $\Deadline(\Zertex_{\Position+k})$ instead of the end depot arrival time $\ArrivalTime(\Route^*)$ defining the planning horizon.

The method inserts $\Zertex$ into $\Memory'(\Vertex\Wertex,\Position)$ if and only if $\Network_\Zertex$ contains a feasible path from $\Zertex_i$ to $\Zertex_j$ traversing arc $\Vertex_\Position\Wertex_{\Position+1}$. Such a path exists if and only if the minimum arrival time at $\Vertex_\Position$ from $\Zertex_i$ is no greater than the maximum feasible departure time from $\Wertex_{\Position+1}$ to $\Zertex_j$. These quantities are computed for every pair of layers $i,j \in \Range{1,\CustCount}$ using time-dependent versions of Dijkstra's algorithm over $\Network_\Zertex$.

The clean-ng method removes unnecessary memory restrictions, enabling additional label dominances while preserving the ng-feasibility status of all tours. Its effectiveness increases after variable fixing/spar\-si\-fi\-ca\-tion followed by time-window shrinking, since tighter arc-position time windows allow further memory entries to be safely removed.

\section{Algorithm Configuration}
\label{app:parametrization}

The following paragraphs describe the configuration of the algorithmic components used in our computational experiments.

\paragraph{Branching.} We process branch-and-bound nodes using best-bound search. At each branching step, we evaluate the $\CustCount/2$ most fractional arcs by solving the restricted $\Master$ after fixing each arc to an integer value. We then select the $\min\{3,k\}$ best candidates for restricted column generation, where $k$ is the minimum integer satisfying
\begin{displaymath}
\LowerBound + (0.01+0.005k)\ArrivalTime(\Route^*) > \ArrivalTime(\Route^*).
\end{displaymath}
Each restricted column generation applies the first three stages of Algorithm~\ref{alg:column generation} to add columns to the restricted $\Master$. Candidates are evaluated using a product rule based on the optimal value of the resulting restricted $\Master$. The primal heuristic is applied in each restricted column generation, and master problems are initialized with columns from previous iterations (Section~\ref{sec:column generation:acceleration}). The limited numbers of branching candidates and column generation stages are motivated by the one-hour time limit, as allocating substantially more computational effort to these procedures provides limited expected benefits.

\paragraph{Solver.} To integrate the solver (Algorithm~\ref{alg:solver}) into the branch-and-bound framework, we modify its stopping condition (Step~\ref{alg:solver:loop start}). Instead of running to completion ($\LowerBound < \ArrivalTime(\Route^*)$), we record the $\Profit$-bounds $\PricingLowerBound_0$, $\PricingLowerBound_1$, and $\PricingLowerBound_2$ at the beginning, after column generation (Step~\ref{alg:solver:CG}), and at the end of each iteration, respectively. The solver stops if either $g_1>0.925g_0$ and $g_2>0.95g_1$, or $g_2>0.9g_0$, where $g_i=\ArrivalTime(\Route^*)-\PricingLowerBound_i$ for $i\in\Range{2}$. Thus, a new iteration is performed only if it achieves sufficient gap reduction, either through a substantial improvement in one phase or through the combined effect of column generation and ng-memory augmentation.

\paragraph{Column Generation.} Each invocation of the column generation method (Algorithm~\ref{alg:column generation}) starts with stage $\Stage_0$, which applies relax-all labeling, followed by four stages $\Stage_1,\ldots,\Stage_4$ using relax-ng labeling with relaxed stabilization. The subsequent stages depend on whether the method is invoked for the first time at the root node of the branch-and-bound tree. In this case, the goal is to obtain strong completion bounds for applying variable fixing as early as possible and avoid repeated pricing on the complete transport network. Therefore, the intermediate exact stages with sparsification are skipped, and stage $\Stage_5$ performs final exact pricing without sparsification. Variable fixing is typically applied immediately after this iteration, although some instances require additional exact pricing iterations on the complete transport network. In subsequent invocations, stages $\Stage_5$ and $\Stage_6$ apply exact labeling with sparsification, while stage $\Stage_7$ performs final exact pricing.

We set $\MaxPricingIterations = 5$ as the minimum number of pricing iterations performed within a stage before moving to the next one. Table~\ref{tab:stages} summarizes the configuration of each stage, where labeling denotes the labeling method, stabilization the stabilization strategy, maxcols the maximum number of tours inserted as columns of $\Master$ after each pricing iteration, $\kappa$ the phase completion criterion parameter, $\SparsificationCost'$ the parameter used to compute the sparsification threshold $\SparsificationCost$ (see below), and $q'$ the parameter used to compute $q$, which controls the frequency of variable fixing (see below).

\begin{table}[htb]
    \centering
    \begin{tabular}{lllrlll}
    \toprule
    Stage               & labeling  & stabilization & maxcols & $\kappa$  & $\SparsificationCost'$ & $q'$ \\\midrule
    $S_0$               & relax-all & none          & 20      & 0         & ---                    & ---  \\
    $S_1$               & relax-ng  & relaxed       & 40      & 0         & 0.05                   & 0.2  \\
    $S_2$               & relax-ng  & relaxed       & 40      & 0         & 0.25                   & 0.2  \\
    $S_3$               & relax-ng  & relaxed       & 40      & 0         & 0.5                    & 0.2  \\
    $S_4$               & relax-ng  & relaxed       & 80      & 0         & 1                      & 0.2  \\
    $S_5$ (first)       & exact     & none          & 100     & 0.95      & ---                    & ---  \\
    $S_5$ (rest)        & exact     & non-relaxed   & 80      & 0.999     & 0.2                    & 0.3  \\
    $S_6$               & exact     & non-relaxed   & 80      & 0.999     & 0.4                    & 0.3  \\
    $S_7$               & exact     & none          & 100     & text      & ---                    & ---  \\\bottomrule
    \end{tabular}
    \caption{Parametrization of the column generation stages}\label{tab:stages}
\end{table}

Regarding $\SparsificationCost$, recall that it replaces $\RedArrivalTime(\Route^*)$ when computing completion values in \eqref{eq:completion:value}. We define it as
\begin{displaymath}
\SparsificationCost = (1-\SparsificationCost')\RedArrivalTime(\Route) + \max\{0,\SparsificationCost'\RedArrivalTime(\Route^*)\},
\end{displaymath}
where $\Route$ is the tour minimizing $\RedArrivalTime$ obtained from the forward labeling before sparsification. When $\RedArrivalTime(\Route^*)\geq 0$, $\SparsificationCost$ is an intermediate threshold between $\RedArrivalTime(\Route^*)$ and the best reduced cost. The $\max$ term handles early stages, where typically $\RedArrivalTime(\Route^*)\ll0$, by preventing the first sparsification step from removing too many arcs and causing a premature stage transition.

For $q'$, we aim to avoid sparsification when its computational cost is not justified by the expected improvement. Therefore, we require the $\Profit$-bound $\PricingLowerBound$ to close a fraction $q'$ of the gap to $\max\{\RMPBound,\ArrivalTime(\Route^*)\}$ between consecutive sparsification invocations, and set
\begin{displaymath}
q = (1-q')\PricingLowerBound + q'\max\{\RMPBound, \ArrivalTime(\Route^*)\}
\end{displaymath}
in Step~\ref{alg:column generation:reinit} of Algorithm~\ref{alg:column generation}. Recall that $\RMPBound$ is the optimal value of the restricted $\Master$; it is typically higher than $\ArrivalTime(\Route^*)$ in early stages and lower in later stages. Additionally, after the initial sparsification, at least $\MaxPricingIterations=5$ pricing iterations are required before applying sparsification again.

Finally, in stage $\Stage_7$, we modify the strategy for setting $\kappa$ to enforce sufficient progress of the $\Profit$-bound $\PricingLowerBound$. Column generation stops when
\begin{displaymath}
\PricingLowerBound \geq 0.85\PricingLowerBound' + 0.15\min\{\RMPBound, \ArrivalTime(\Route^*)\},
\end{displaymath}
where $\PricingLowerBound'$ is the last $\Profit$-bound obtained before starting column generation. Note that $\min\{\RMPBound,\ArrivalTime(\Route^*)\}$ is an upper bound on the $\Profit$-bound, even when $\Profit$ corresponds to the optimal dual solution of $\Master$.

\paragraph{Initialization of $\Master$.} We preserve at least half of the columns of $\Master$ after ng-memory augmentation. Columns corresponding to tours that are no longer ng-feasible receive a penalty of $\ArrivalTime(\Route^*)-\PricingLowerBound_1$, where $\PricingLowerBound_1$ is the $\Profit$-bound obtained at the end of the column generation phase.

\paragraph{Ng-memory initialization.} The initial ng-memory structure emulates a vertex-based ng-memory in which each vertex stores its five closest customers. Thus, for $\Vertex,\Wertex,\Zertex,\Position \in \Range{1,\CustCount}$, we insert $\Zertex$ into $\Memory(\Vertex\Wertex,\Position)$ if $\Vertex\neq\Wertex$ and both are among the five customers closest to $\Zertex$.

\paragraph{Ng-memory augmentation.} During the first three applications of the cycle-forbidding method in the forward direction, the threshold $\kappa$ controlling the stopping condition is set to $\infty$, so the pricing problem is solved at least four times in each direction. For the $i$-th forward pricing problem with $i>3$, we set $\kappa=0.93\PricingLowerBound_{i-3}$, where $\PricingLowerBound_j$ is the $\Profit$-bound obtained after the $j$-th iteration. Thus, the method continues only if the improvement in the $\Profit$-bound is sufficient. The set of tours computed in Step~\ref{alg:MA:tours}, whose cycles are forbidden in Step~\ref{alg:MA:forbid}, contains at most $60\CustCount$ tours.

\paragraph{Exact search.} We consider four equally spaced candidate thresholds $t_0,\ldots,t_3$. For each $t_i$, the computational budget is limited to $\max\{10^6,k\}$ non-discarded labels, where $k$ is the number of non-dominated labels in the last pricing problem solved in the preceding ng-memory augmentation.

\paragraph{Primal heuristic.} We set the maximum number of non-improving iterations of the destroy-and-repair heuristic to $\CustCount/2$. In the destroy step, the number of removed customers is selected uniformly from $\lfloor\CustCount/20\rfloor$ to $\lfloor\CustCount/10\rfloor$, and the starting position of the removed subsequence is chosen uniformly at random. For local search, we use a first-improvement strategy and stop when no improving move is found.

\paragraph{Primal heuristic time budgets.} At the end of each column generation phase (Step~\ref{alg:column generation:primal} of Algorithm~\ref{alg:column generation}), we run the primal heuristic with a time budget equal to $\Time/5$, where $\Time$ is the phase duration. Similarly, after each forward pricing problem in the MA method (Step~\ref{alg:MA:primal} of Algorithm~\ref{alg:MA}), we allocate a time budget of $\Time/5$, where $\Time$ is the duration of the corresponding labeling method.  At most a quarter of this time is spent converting non-elementary tours found in the current phase into elementary tours, processing them in increasing order of $\ArrivalTime$. The resulting tours are stored in a pool, from which those with smallest $\ArrivalTime$ are selected for the destroy-and-repair heuristic. Elementary tours not selected remain in the pool and are considered in subsequent executions of the primal heuristic, and the pool is emptied at the end of each iteration of Algorithm~\ref{alg:solver}.

\section{Extended Experimental Results}
\label{app:results}

Tables \ref{tab:app:1}--\ref{tab:app:3} report aggregated results for benchmark instances considered in the main text but not reported there.

\begin{table}[htb]
    \centering\small
    \begin{tabular}{lcc ccc ccccc}
        \toprule
         &   &   & \multicolumn{2}{c}{\Lera} && \multicolumn{5}{c}{\BP} \\\cmidrule{4-5}\cmidrule{7-11}
                   &    \#   & $\CustCount$  & $s$ & $t$     && $s$ & $t$   & $t_{\max}$ & $\times_t$  & $\times_m$ \\\midrule
      \Adamo       &   180   & 15            & 180 & 38      && 180 & 0     & 1          & 0           & 1          \\
                   &   180   & 25            & 164 & 220     && 180 & 2     & 10         & 3           & 13         \\
                   &   180   & 35            & 145 & 2617    && 180 & 14    & 34         & 18          & 42         \\
                   &   180   & 45            & --- & ---     && 180 & 66    & 220        & ---         & ---        \\
                   &   180   & 55            & --- & ---     && 176 & 369   & 2986       & ---         & ---        \\\midrule
      \RifkiNoTW   &    30   & 10            & --- & ---     && 30  & 0     & 0          & ---         & ---        \\
                   &    30   & 20            & --- & ---     && 30  & 1     & 1          & ---         & ---        \\
    \end{tabular}
    \caption{Missing results for instances of the \TDTSP{} (no time windows).}\label{tab:app:1}
\end{table}

\begin{table}[htb]
    \centering\small
    \begin{tabular}{lccc ccc ccc ccccc}
        \toprule
         &   &   &    & \multicolumn{2}{c}{\Lera} && \multicolumn{2}{c}{\Fontaine} && \multicolumn{5}{c}{\BP} \\\cmidrule{5-6}\cmidrule{8-9}\cmidrule{11-15}
                   &    \#   & $\CustCount$  & $\TW$ & $s$ & $t$     && $s$      & $t$     && $s$ & $t$   & $t_{\max}$ & $\times_t$  & $\times_m$ \\\midrule
      \Arigliano   &    300  & 15            & 0     & 300 & 5       && 300      & 0       && 300 & 0     & 0          & 0           & 1          \\
                   &    300  & 20            & 0     & 300 & 198     && 300      & 1       && 300 & 0     & 1          & 1           & 4          \\\midrule
      \Rifki       &    150  & 10            & 0     & --- & ---     && ---      & ---     && 150 & 0     & 0          & 0           & 0          \\
                   &    150  & 20            & 0     & 148 & 363     && 150      & 0       && 150 & 0     & 2          & 1           & 4          \\\midrule
      \RifkiConst  &    60   & 20            & 0     & 60  & 59      && 60       & 0       && 60  & 0     & 1          & 1           & 1          \\\midrule
      \Arigliano   &    300  & 15            & 0.25  & 300 & 3       && 300      & 0       && 300 & 0     & 0          & 0           & 1          \\
                   &    300  & 20            & 0.25  & 300 & 87      && 300      & 0       && 300 & 0     & 2          & 1           & 6          \\\midrule
      \Rifki       &    150  & 10            & 0.25  & --- & ---     && ---      & ---     && 150 & 0     & 0          & 0           & 0          \\
                   &    150  & 20            & 0.25  & 150 & 89      && 150      & 0       && 150 & 0     & 1          & 1           & 2          \\\midrule
      \RifkiConst  &    60   & 20            & 0.25  & 60  & 16      && 60       & 0       && 60  & 0     & 0          & 0           & 1          \\\midrule
      \Arigliano   &    300  & 15            & 0.5   & 300 & 2       && 300      & 0       && 300 & 0     & 0          & 0           & 0          \\
                   &    300  & 20            & 0.5   & 300 & 19      && 300      & 0       && 300 & 0     & 1          & 1           & 2          \\\midrule
      \Rifki       &    150  & 10            & 0.5   & --- & ---     && ---      & ---     && 150 & 0     & 0          & 0           & 0          \\
                   &    150  & 20            & 0.5   & 150 & 15      && 150      & 0       && 150 & 0     & 1          & 1           & 1          \\\midrule
      \RifkiConst  &    60   & 20            & 0.5   & 60  & 5       && 60       & 0       && 60  & 0     & 0          & 0           & 1          \\\bottomrule
  \end{tabular}
  \caption{Missing results for instances of the \TDTSPTW{} with loose and moderate time windows.}\label{tab:app:2}
\end{table}

\begin{table}[htb]
    \centering\small
    \begin{tabular}{lccc ccc ccc ccccc}
        \toprule
         &   &   &    & \multicolumn{2}{c}{\Lera} && \multicolumn{2}{c}{\Fontaine} && \multicolumn{5}{c}{\BP} \\\cmidrule{5-6}\cmidrule{8-9}\cmidrule{11-15}
                   &    \#   & $\CustCount$  & $w$/$\TW$ & $s$ & $t$     && $s$      & $t$     && $s$ & $t$   & $t_{\max}$ & $\times_t$  & $\times_m$ \\\midrule
      \Vu          &    120  & 60            & $\leq 80$ & 120 & 1       && 120      & 0       && 120 & 3     & 5          & 7           & 12         \\
                   &    40   & 60            & 100       & 40  & 8       && 40       & 0       && 40  & 4     & 6          & 10          & 14         \\
                   &    40   & 60            & 120       & 40  & 26      && 40       & 0       && 40  & 5     & 9          & 14          & 23         \\
                   &    40   & 60            & 150       & 40  & 155     && 40       & 0       && 40  & 7     & 14         & 17          & 36         \\\midrule
                   &    120  & 80            & $\leq 80$ & 120 & 8       && 120      & 0       && 120 & 10    & 19         & 25          & 47         \\
                   &    40   & 80            & 100       & 40  & 53      && 40       & 0       && 40  & 17    & 31         & 42          & 78         \\
                   &    40   & 80            & 120       & 40  & 97      && 40       & 0       && 40  & 20    & 47         & 49          & 118        \\
                   &    40   & 80            & 150       & 40  & 193     && 40       & 2       && 40  & 32    & 260        & 79          & 653        \\\midrule
      \Arigliano   &    300  & 15            & 100       & 300 & 0       && 300      & 0       && 300 & 0     & 0          & 0           & 0          \\
                   &    300  & 20            & 100       & 300 & 0       && 300      & 0       && 300 & 0     & 0          & 0           & 0          \\
                   &    300  & 30            & 100       & 300 & 0       && 300      & 0       && 300 & 0     & 1          & 1           & 1          \\
                   &    300  & 40            & 100       & 300 & 0       && 300      & 0       && 300 & 1     & 2          & 2           & 4          \\\midrule
      \Rifki       &    150  & 10            & 100       & --- & ---     && ---      & ---     && 150 & 0     & 0          & 0           & 0          \\
                   &    150  & 20            & 100       & 150 & 0       && 150      & 0       && 150 & 0     & 0          & 0           & 0          \\
                   &    150  & 30            & 100       & 150 & 0       && 150      & 0       && 150 & 0     & 1          & 1           & 2          \\
                   &    150  & 40            & 100       & 150 & 0       && 150      & 0       && 150 & 1     & 2          & 2           & 5          \\
                   &    150  & 50            & 100       & --- & ---     && ---      & ---     && 150 & 2     & 5          & 5           & 14         \\
                   &    150  & 60            & 100       & --- & ---     && ---      & ---     && 150 & 4     & 7          & 11          & 17         \\\midrule
      \RifkiConst  &    60   & 20            & 100       & 60  & 0       && 60       & 0       && 60  & 0     & 0          & 0           & 0          \\
                   &    60   & 30            & 100       & 60  & 0       && 60       & 0       && 60  & 0     & 0          & 0           & 0          \\
                   &    60   & 40            & 100       & 60  & 0       && 60       & 0       && 60  & 0     & 1          & 1           & 2          \\\bottomrule
  \end{tabular}
  \caption{Missing results for instances of the \TDTSPTW{} with tight time windows.}\label{tab:app:3}
\end{table}

\end{document}